\documentclass[twocolumn]{revtex4-2}
\usepackage[dvipdfmx]{graphicx}
\usepackage[T1]{fontenc}
\usepackage{lmodern}
\usepackage{here}
\usepackage{amsmath}
\usepackage{amssymb}
\usepackage{bm}
\usepackage{color}
\begin{document}
\allowdisplaybreaks[4]

\title{Stability of supercurrents in a superfluid phase of spin-1 bosons in an optical lattice}
\author{Shion Yamashika$^{1}$, Ryosuke Yoshii$^{2}$, and Shunji Tsuchiya$^{1}$}
\affiliation{$^{1}$Department of Physics, Chuo University, 1-13-27 Kasuga, Tokyo 112-8551, Japan\\
 $^{2}$Center for Liberal Arts and Sciences, Sanyo-Onoda City University, 1-1-1 Daigaku-Dori, Sanyo-Onoda, Yamaguchi 756-0884, Japan}
\date{\today}
\begin{abstract}
  We study collective modes and superfluidity of spin-1 bosons with antiferromagnetic interactions in an optical lattice based on the time-dependent Ginzburg-Landau (TDGL) equation derived from the spin-1 Bose-Hubbard model. Specifically, we examine the stability of supercurrents in the polar phase in the vicinity of the Mott insulating phase with even filling factors. Solving the linearized TDGL equation, we obtain gapless spin-nematic modes and gapful spin-wave modes in the polar phase that arise due to the breaking of $S^2$ symmetry in spin space. Supercurrents exhibit dynamical instabilities induced by growing collective modes. In contrast to the second-order phase transition, the critical momentum for mass currents is finite at the phase boundary of the first-order superfluid-Mott insulator (SF-MI) phase transition. Furthermore, the critical momentum remains finite throughout the metastable SF phase and approaches zero towards the phase boundary, at which the metastable SF state disappears. We also study the stability of spin currents motivated by recent experiments for spinor gases. The critical momentum for spin currents is found to be zero, where a spin-nematic mode causes the dynamical instability. We investigate the origin of the zero critical momentum for spin currents and find it attributed to the fact that the polar state becomes energetically unstable even in the presence of an infinitesimal spin current. We discuss implications of the zero critical momentum for spin currents for the stability of the polar state.
\end{abstract}
\maketitle
\section{Introduction}
Ultracold atoms in an optical lattice have opened a versatile research field that lies at the interface of condensed matter physics, statistical physics, and atomic, molecular, and optical physics \cite{lewenstein07,lewenstein12}. Spinor bosons in an optical lattice provide with an ideal platform for studying quantum magnetism, quantum phase transitions, and non equilibrium quantum dynamics \cite{garcia04,imambekov04,widera05,yamashita07,shinozaki13,stamper-kurn13,fujimoto19}. In particular, the superfluid-Mott insulator (SF-MI) phase transition of spin-1 bosons in an optical lattice has attracted much attention. The rich structure of the phase diagram that arises from the interplay between strong correlations and spin degrees of freedom has been a main focus of intense theoretical studies. It exhibits many interesting features including the parity effect of the MI phase \cite{demler02,tsuchiya04,rizzi05,lacki11}, the first-order SF-MI transition \cite{kimura05,krutitsky05,batrouni05,yamamoto13}, and the spin-nematic order in the MI phase \cite{yip03,imambekov03,zhou03,snoek04,rizzi05,apaja06,bernier06,deforgesdeparny13,alavani18,deforgesdeparny18}.
On the other hand, recent progress of experimental studies have made possible to observe the signature of the first-order phase transition \cite{liu16} and quantum critical dynamics \cite{austin_2021} in this system.
\par
One of the remarkable features of bosonic superfluids in an optical lattice is dynamical instabilities of supercurrents \cite{burger01,cataliotti03,fallani04,desarlo05,mun07,ferris08, niu01,auerbach02,wu03,polkovnikov05,altman05}.  
In fact, this phenomenon is considered to be deeply related with the SF-MI phase transition.
It has been predicted for spineless bosons that the critical superfluid velocity approaches zero at the second-order SF-MI phase transition reflecting the diverging healing length \cite{altman05,polkovnikov05} and later it was experimentally confirmed \cite{mun07}.
Given the rich physics associated with the SF-MI phase transition and the recent experimental progress, therefore, it is of particular interest to investigate dynamical instabilities of spin-1 bosons in an optical lattice. 
Recently, the critical momentum for mass currents and the stability phase diagram have been calculated numerically using the dynamical Gutzwiller approximation \cite{asaoka16}. The critical momentum in the vicinity of the first-order SF-MI phase transition, however, has not been fully investigated.
Meanwhile, critical dynamics of spin currents has been observed in an antiferromagnetic spin-1 Bose-Einstein condensate \cite{kim17}. In this experiment, spin currents were induced by counter flow of two spin components. 
This experiment inspires us to investigate the stability of spin currents in an optical lattice.
\par
In this paper, we study spin-1 bosons with antiferromagnetic interactions in an optical lattice.
Specifically, we examine the stability of supercurrents in the polar phase in the vicinity of the MI phase with even filling factors
based on the time-dependent Ginzburg-Landau (TDGL) equation, which we derive from the spin-1 Bose-Hubbard model.
Our focus is on the superfluidity in the vicinity of the first-order phase transition.
We calculate the critical momentum for mass currents and find that it has a finite value not only 
at the first-order SF-MI phase boundary, but throughout the metastable SF phase, in contrast to the second-order case.
We also study the stability of spin currents motivated by the recent experiments for spinor gases \cite{kim17}.
The critical momentum for spin currents is found to be zero. We clarify the origin of the instability of spin currents and briefly discuss about its implications for the stability of the polar phase.
\par
The organization of this paper is as follows: In Sec.~II, we introduce the spin-1 Bose-Hubbard model and calculate the metastable phase diagram within the perturbative mean-field theory.
In Sec.~III, we study collective modes in the SF phase using the TDGL equation. In Sec.~IV, we study the stability of supercurrents and calculate the critical momenta for mass and spin currents. In Sec.~V, we investigate the origin of the instability of spin currents and discuss the stability of the polar phase. Finally, we summarize our results in Sec.~VI. Details of the calculations in the perturbative mean-field theory are given in Appendix A and B. The derivations of the TDGL equation and the conservation laws are given in Appendix C and D, respectively. We set $\hbar=k_{\rm B}=1$ throughout this paper. 
\section{spin-1 Bose-Hubbard model and SF-MI phase transition}
We consider spin-1 bosons trapped in a $d$-dimensional cubic optical lattice at zero temperature. In this paper, we neglect effect of a harmonic trapping potential for simplicity. If the lattice potential is sufficiently deep, the system is well described by the spin-1 Bose-Hubbard model \cite{tsuchiya04,imambekov03}:
\begin{eqnarray}
\hat {H}&=&-t\sum_{\langle i,j\rangle,\alpha}(\hat b^\dagger_{i\alpha}\hat b_{j\alpha}+\mathrm{h.c.})-\mu\sum_i \hat n_i\nonumber\\
&+&\frac{U_0}{2}\sum_i \hat n_i(\hat{n}_i-1)+\frac{U_2}{2}\sum_{i}(\hat{\bm{S}}_i^2-2\hat n_i),\label{BHM}
\end{eqnarray}
where $\hat{b}_{i\alpha}\,(\hat{b}_{i\alpha}^\dagger)$ is the annihilation (creation) operator for an atom in the hyperfine state $|F=1,m=\alpha\rangle\,(\alpha=1,0,-1)$ at site $i$. Here, $\langle i,j \rangle$ denotes a summation over nearest-neighbor sites, $t$ the hopping matrix element, $\mu$ the chemical potential, and $U_0$ and $U_2$ the on-site spin-independent and spin-dependent interactions, respectively. In this paper, we assume $U_2>0$, i.e., an antiferromagnetic interaction. $U_2/U_0$ is fixed for each atomic species. For example, $U_2/U_0=0.04$ for $^{23}{\rm Na}$ \cite{tsuchiya04}. In this paper, we set $U_2/U_0$ optimal values for validity of the theory. The operators $\hat{n}_i=\sum_{\alpha}\hat{b}_{i\alpha}^\dagger\hat{b}_{i\alpha}$, and $\hat{\bm{S}}_i=\sum_{\alpha,\beta}\hat{b}_{i\alpha}^\dagger\bm{F}_{\alpha\beta}\hat{b}_{i\beta}$ represent the number of particles and the spin at site $i$, respectively. Here, $\bm{F}$ denotes the spin-1 matrices
\begin{eqnarray}
F_x&=&\frac{1}{\sqrt{2}}
\left[
\begin{array}{ccc}
0&1&0\\
1&0&1\\
0&1&0
\end{array}
\right],\ 
F_y=\frac{i}{\sqrt{2}}
\left[
\begin{array}{ccc}
0&-1&0\\
1&0&-1\\
0&1&0
\end{array}
\right],\nonumber \\
F_z&=&
\left[
\begin{array}{ccc}
1&0&0\\
0&0&0\\
0&0&-1
\end{array}
\right].
\end{eqnarray}\\
\indent In the following subsections, we briefly review the perturbative mean-field theory developed in Ref.~\cite{tsuchiya04} and then extend it to describe the first-order SF-MI phase transition as well as the metastable SF and MI phases.
\subsection{Mott insulating phase}\label{sec:mott_insulating_phase}
We first study the MI state in the limit of $t=0$, where the Hamiltonian (\ref{BHM}) can be written as  
\begin{eqnarray}
\hat{H}&=&\sum_i\hat{H}^0_i,\\
\hat{H}^0_i&=&-\mu\hat{n}_i+\frac{U_0}{2}\hat{n}_i(\hat{n}_i-1)+\frac{U_2}{2}(\hat{\bm{S}}_i^2-2\hat{n}_i).\label{H0}
\end{eqnarray}
Note that $\hat{\bm{S}}^2_i$, $\hat{S}_{iz}$ and $\hat{n}_i$ commute with each other. The simultaneous eigenstates for these operators $|S_i,m_i,n_i\rangle\  (-S_i\leq m_i\leq S_i)$ satisfy
\begin{eqnarray}
\hat{\bm{S}}_i^2|S_i,m_i,n_i\rangle&=&S_i(S_i+1)|S_i,m_i,n_i\rangle,\label{S^2}\\
\hat{S}_{iz}|S_i,m_i,n_i\rangle&=&m_i|S_i,m_i,n_i\rangle,\label{S_z}\\
\hat{n}_i|S_i,m_i,n_i\rangle&=&n_i|S_i,m_i,n_i\rangle\label{n}.
\end{eqnarray}
$|S_i,m_i,n_i\rangle$ is thus an energy eigenstate of $\hat{H}^0_i$:
\begin{eqnarray}
\hat{H}^0_i|S_i,m_i,n_i\rangle=E^0(S_i,n_i)|S_i,m_i,n_i\rangle,
\end{eqnarray}
where the energy eigenvalue $E^0(S_i,n_i)$ is given by
\begin{eqnarray}
E^0(S_i,n_i)&=&-\mu n_i+\frac{U_0}{2}n_i(n_i-1)\nonumber \\
&&+\frac{U_2}{2}[S_i(S_i+1)-2n_i].
\end{eqnarray}
$S_i$ takes the minimum value in the ground state due to $U_2>0$.\\
\indent Since the orbital part of the wave function in a single lattice site is symmetric under permutation of any two atoms, the spin part has to be also symmetric under permutation of atoms due to Bose statistics. As a result, $S_i$ is even $(S_i=0,2,4,...,n_i)$ for even $n_i$, while $S_i$ is odd $(S_i=1,3,5,...,n_i)$ for odd $n_i$ \cite{wu96}.
The ground state of the single-site Hamiltonian (\ref{H0}) is thus $|0,0,n_i\rangle$ for even filling factors, while it is $|1,m_i,n_i\rangle\, (m_i=1,0,-1)$ for odd filling factors that is triply degenerate. For even filling factors, the ground state $|0,0,n_i\rangle$ is a spin-singlet insulator, in which all atoms form spin-singlet pairs \cite{tsuchiya04} as
\begin{eqnarray}
|0,0,n_i\rangle=\frac{1}{\sqrt{f(\frac{n_i}{2};0)}}(\hat{\Theta}_i^\dagger)^{\frac{n_i}{2}}|\mathrm{vac}\rangle,\label{spin-singlet-mi}
\end{eqnarray}
where $\hat{\Theta}^\dagger_i=(\hat{b}^\dagger_{i0})^2-2\hat{b}^\dagger_{i1}\hat{b}^\dagger_{i-1}$ is the creation operator of a spin-singlet pair. The normalization factor is given by \cite{yip00}
\begin{eqnarray}
f(Q;S)=S!Q!2^Q\frac{(2Q+2S+1)!!}{(2S+1)!!}.
\end{eqnarray}
For odd filling factors, there remains a single atom that cannot form a spin-singlet pair in the ground state as
\begin{eqnarray}
|1,m_i,n_i\rangle =\frac{1}{\sqrt{f(\frac{n_i-1}{2};1)}}\hat{b}^\dagger_{i\,m_i}(\hat{\Theta}^\dagger_i)^{\frac{n_i-1}{2}}|\mathrm{vac}\rangle. \label{odd}
\end{eqnarray} 

\subsection{Perturbative mean-field theory}\label{sec:perturbaticve_mean-field_theory}
We employ the perturbative mean-field theory to study the SF-MI phase transition \cite{tsuchiya04}. We introduce the superfluid order parameter $\psi_\alpha \equiv \langle \hat{b}_{i\alpha}\rangle\equiv \sqrt{n_{\rm c}}\zeta_\alpha$, where $n_{\rm c}=\sum_{\alpha=\pm1,0} |\psi_\alpha|^2$ is the number of condensate atoms per site and $\zeta_\alpha$ is the normalized spinor: $\sum_{\alpha=\pm1,0}\zeta^*_\alpha\zeta_\alpha=1$. Linearizing the hopping term with respect to fluctuation $\delta \hat{b}_{i\alpha}=\hat{b}_{i\alpha}-\psi_{\alpha}$, we obtain
\begin{eqnarray}
&&-t\sum_{\langle i,j\rangle,\alpha}(\hat{b}^\dagger_{i\alpha}\hat{b}_{j\alpha}+\mathrm{h.c.})\nonumber\\
&&\sim -zt\sum_{i,\alpha}[ (\psi_{\alpha}^*\hat{b}_{i\alpha}+\psi_\alpha\hat{b}_{i\alpha}^\dagger)]+ztN_{\rm s}\sum_\alpha\psi_\alpha^*\psi_\alpha,\label{hopping}
\end{eqnarray}
where $z= 2d$ denotes the number of nearest-neighbor sites and $N_{\rm s}$ the total number of lattice sites. The Hamiltonian (\ref{BHM}) thus reduces to 
\begin{eqnarray}
\hat{H}&=&\sum_{i}\hat{H}^\mathrm{mf}_i,\label{eff_H}\\
\hat{H}^\mathrm{mf}_i&=&\hat{H}_i^0+\hat{V}_i+zt\sum_\alpha\psi_\alpha^*\psi_\alpha,\label{H_i}\\
\hat V_i&=&-zt\sum_{\alpha}(\psi_\alpha\hat{b}_{i\alpha}^\dagger+\psi_\alpha^*\hat{b}_{i\alpha}).
\end{eqnarray}
Here, $\hat{V}_i$ describes transfer of atoms between the $i$-th site and the condensate $\psi_\alpha$. We treat $\hat{V}_i$ as a perturbation assuming small {\it t}. We omit the site index in the rest of this section. \\
\indent We focus on the MI phase with even filling factors and the SF phase around it in this paper. In this case, since the MI phase has a unique ground state (\ref{spin-singlet-mi}), the standard perturbative mean-field theory is applicable. For odd filling factors, extensions are required for the degenerate ground state (\ref{odd}).\\
\indent To describe the first-order SF-MI phase transition predicted in Ref.\,\cite{kimura05}, it is necessary to expand the ground-state energy up to sixth-order. Applying the standard perturbation theory, the ground-state energy per site can be calculated as
\begin{eqnarray}
 E&=&E^0(0,n)+C_2n_{\rm c}+C_4n_{\rm c}^2+C_4'n_{\rm c}^2\langle \bm{F}\rangle^2\nonumber\\
 &&+C_6n_{\rm c}^3+C_6'n_{\rm c}^3\langle \bm{F}\rangle^2,\label{free}
\end{eqnarray}
where $\vec{\zeta}=(\zeta_1,\zeta_0,\zeta_{-1})^T$ and $\langle\bm{F}\rangle=\sum_{\alpha,\beta}\zeta_{\alpha}^*\bm{F}_{\alpha\beta}\zeta_{\beta}$.
The details of the calculations and the explicit forms of $C_2$, $C_4$, $C_4'$, $C_6$, and $C_6'$ are given in Appendix B. The fourth-order term with $C_4'>0$ lifts the degeneracy in spin-space and the polar state is realized in the SF phase \cite{tsuchiya04}: $\langle \bm{F}\rangle^2$ takes its minimum in the ground state, i.e., $\langle \bm{F}\rangle=\bm 0$.\\
\indent Since all spinors are related to each other by gauge transformation $e^{i\theta}$ and spin rotations $R(\alpha,\beta,\gamma)=e^{-i\alpha F_z}e^{-i\beta F_y}e^{-i\gamma F_z}$, where $(\alpha,\beta,\gamma)$ are the Euler angles, the general form of the order parameter for the polar state is given as
\cite{ho98}
\begin{eqnarray}
\vec \zeta=e^{i\theta}R(\alpha,\beta,\gamma)\left[
\begin{array}{c}
0\\
1\\
0
\end{array}
\right]
=e^{i\theta}\left[
\begin{array}{c}
-\frac{e^{-i\alpha}}{\sqrt{2}}\sin\beta\\
\cos\beta\\
\frac{e^{i\alpha}}{\sqrt 2}\sin\beta
\end{array}
\right].\label{polarst}
\end{eqnarray}
Since $\vec{\zeta}$ in Eq.\,(\ref{polarst}) is invariant under shift of $\gamma$ as well as the discrete transformation $(\beta,\theta)\rightarrow(\beta+\pi,\theta+\pi)$, the polar state has the symmetry characterized by the isotropy group $H={\rm U}(1)_\gamma\otimes ({\bf Z}_2)_{\beta,\theta}$ \cite{stamper-kurn13}.  The symmetry of the spin-singlet state in Eq.\,(\ref{spin-singlet-mi}) is characterized by the isotropy group $G={\rm U}(1)_\theta\otimes {\rm SO}(3)_{\alpha,\beta,\gamma}$. The order parameter manifold for the polar state is thus given by \cite{zhou01}
\begin{eqnarray}
M=G/H
&=& \frac{{\rm U}(1)_\theta\otimes {\rm SO}(3)_{\alpha,\beta,\gamma}}{{\rm U}(1)_\gamma \otimes({\mathbf Z}_2)_{\beta,\theta}}\nonumber \\
&=&\frac{{\rm U}(1)_\theta\otimes S^2_{\alpha,\beta}}{({\mathbf Z}_2)_{\beta,\theta}}.\label{symmetry}
\end{eqnarray}
\indent It is convenient to introduce the $d$ vector defined as \cite{machida98} 
\begin{eqnarray}
\bm{d}
=\left[
\begin{array}{c}
d_x\\
d_y\\
d_z
\end{array}
\right]
=
\left[
\begin{array}{c}
\frac{1}{\sqrt{2}}(-\psi_1+\psi_{-1})\\
\frac{1}{\sqrt{2}i}(\psi_1+\psi_{-1})\\
\psi_0
\end{array}
\right]. \label{d-vector}
\end{eqnarray}
Note that the $d$ vector transforms as a vector under rotations in spin-space. From Eq.\,(\ref{polarst}), the polar state is characterized by the $d$ vector 
\begin{eqnarray}
\bm{d}=\sqrt{n_{\rm c}}e^{i\theta}
\left[
\begin{array}{c}
\sin\beta\cos\alpha\\
\sin\beta\sin\alpha\\
\cos\beta
\end{array}
\right],
\end{eqnarray}
where $\alpha$ and $\beta$, respectively, represent the polar and azimuthal angles that parametrize the surface of the unit sphere $S^2$.
Due to the invariance of $\vec{\zeta}$ under $(\beta,\theta)\rightarrow (\beta +\pi,\theta+\pi)$, $\bm{d}$ specifies a preferred axis, not a preferred direction along that axis.
\subsection{Second-order SF-MI phase transition}\label{sec:2nd}
\begin{figure}
\includegraphics[clip,width=220pt]{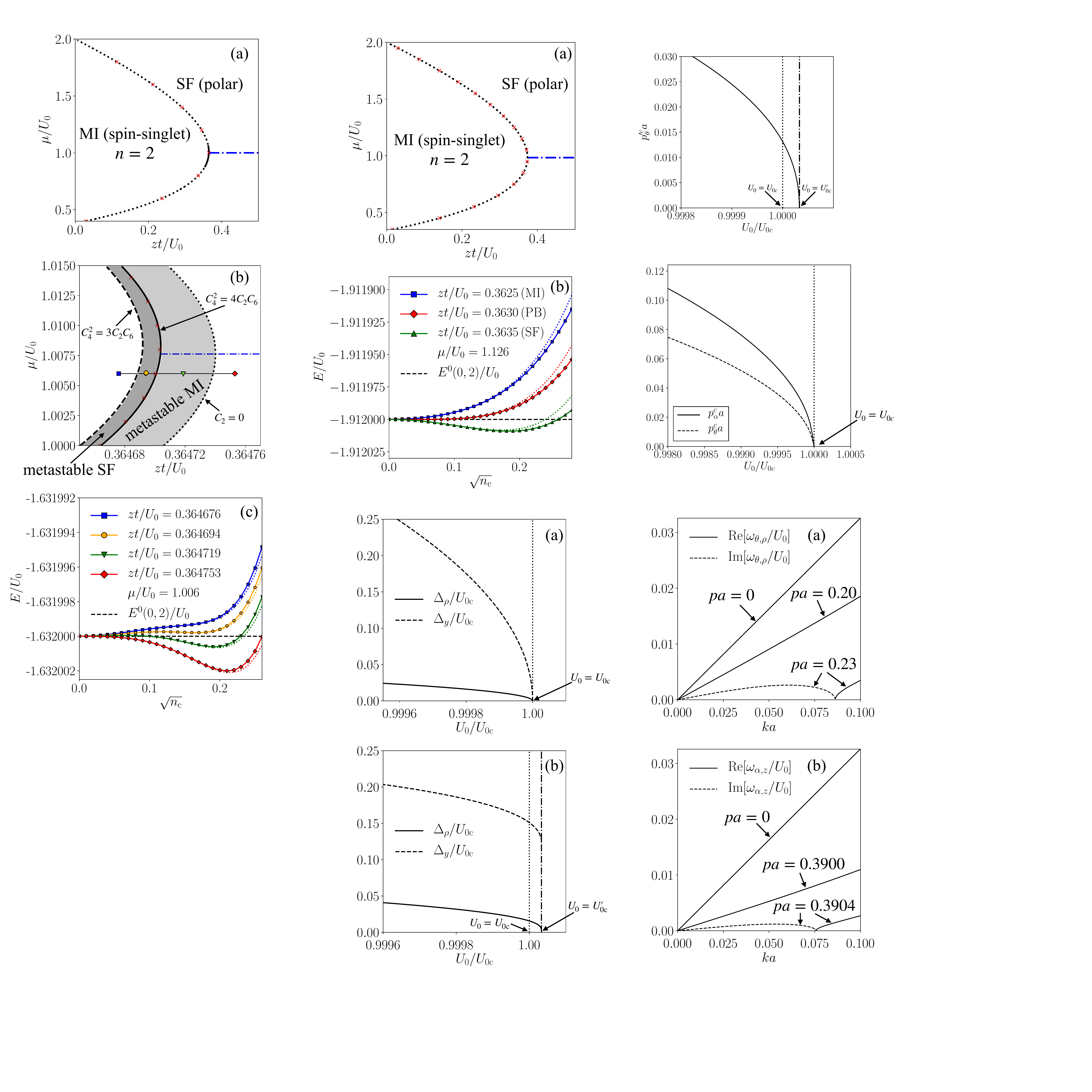}
\caption{(Color online) (a) Phase diagram of the second-order SF-MI phase transition. We set $U_2/U_0=0.33$. The dotted line represents the phase boundary between the spin-singlet insulator with $n=2$ and the polar state determined by the perturbative mean-field theory. The red crosses represent the phase boundary obtained by numerically diagonalizing Eq.\,(\ref{H_i}). The blue dash-dotted line indicates the particle-hole symmetry line determined by $K=0$ (see Appendix D). 
(b) Ground state energy as a function of $\sqrt{n_{\rm c}}$ in the MI phase (squares), the SF phase (triangles), and at the phase boundary (PB) (diamonds). The solid lines plot Eq.\,(\ref{free_4}). The dotted lines are obtained by diagonalizing Eq.\,(\ref{H_i}).}\label{phase_diagram_1}
\end{figure}
In the case of the second-order SF-MI phase transition, where $U_2/U_0>0.32$ \cite{kimura05}, it is enough to expand the ground-state energy up to fourth order as
\begin{eqnarray}
E=E^0(0,n)+C_2n_{\rm c}+C_4n_{\rm c}^2.\label{free_4}
\end{eqnarray}
Note that we have assumed the polar state and set $\langle \bm{F}\rangle=\bm 0$ in Eq.\,(\ref{free_4}). The right-hand side of Eq.\,(\ref{free_4}) is plotted as a function of $\sqrt{n_{\rm c}}$ in Fig.\,\ref{phase_diagram_1}\,(b). From the standard Landau theory for second-order phase transitions \cite{landau80}, the phase boundary between the MI and SF phases is determined by the condition $C_2=0$. Figure\,\ref{phase_diagram_1}\,(a) shows the resulting phase diagram \cite{tsuchiya04}.\\
\indent The ground-state energy can be calculated by numerically diagonalizing Eq.\,(\ref{H_i}), as shown in Fig.\,\ref{phase_diagram_1}\,(b). The phase boundary can be determined by $n_{\rm c}$ that minimizes the numerically calculated ground-state energy. It precisely agrees with that of the perturbative mean-field theory, as shown in Fig.\,\ref{phase_diagram_1}\,(a).

\subsection{First-order SF-MI phase transition}
\begin{figure}%
\includegraphics[clip,width=210pt]{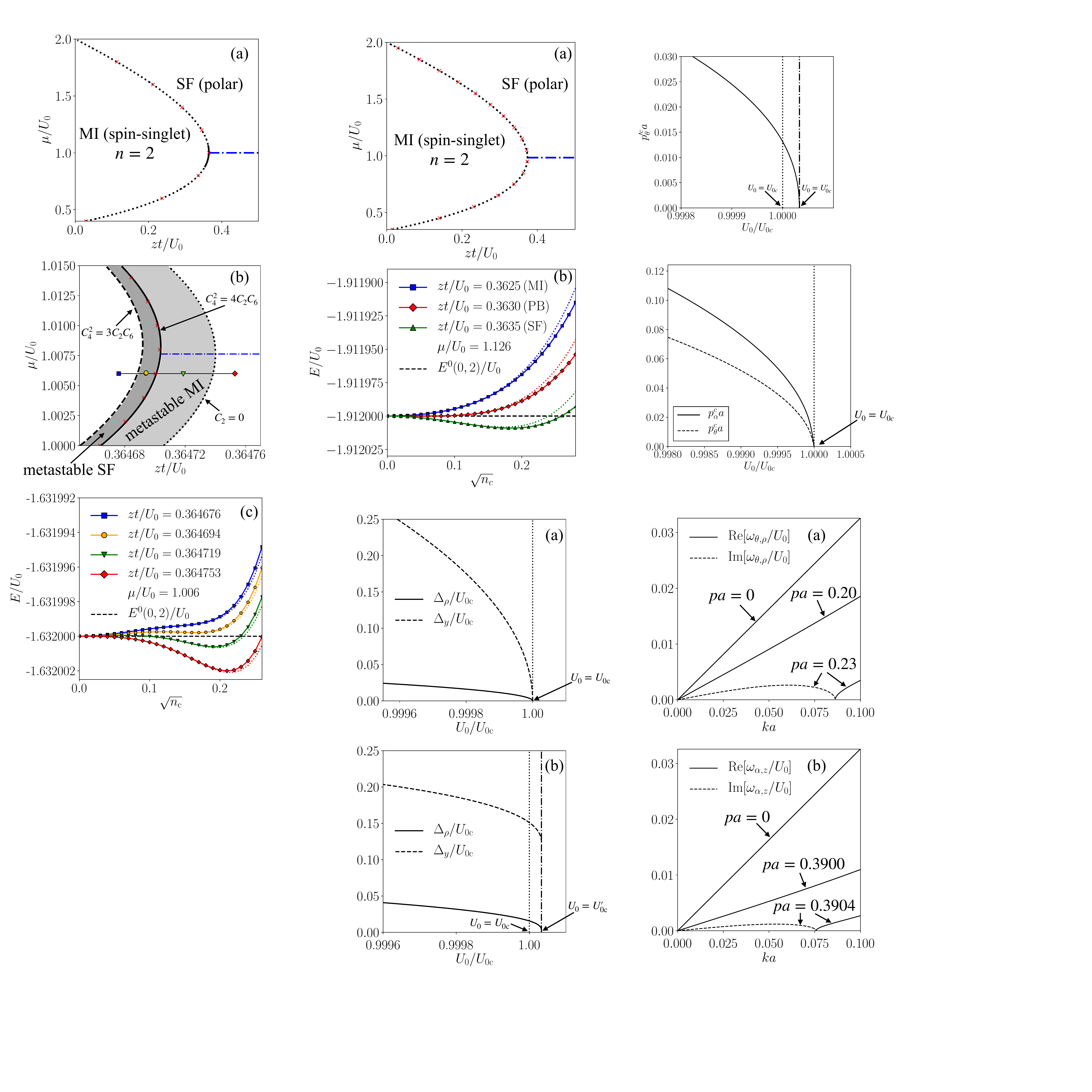}
\caption{(Color online) (a) Phase diagram of the first-order SF-MI phase transition. We set $U_2/U_0=0.31$. The dotted line represents the phase boundary between the spin-singlet insulator with $n=2$ and the polar state determined by the perturbative mean-field theory. The red crosses represent the phase boundary obtained by numerically diagonalizing Eq.\,(\ref{H_i}). The blue dash-dotted line indicates the particle-hole symmetry line determined by $K=0$ (see Appendix D). The solid black line indicates the phase boundary of the first-order phase transition.
(b) Magnification of the tip of the Mott lobe in (a).
The solid  line is the phase boundary.
The metastable SF (MI) state exists in the thick (thin) gray region, in which the condition $3C_2C_6<C_4^2<4C_2C_6$ $(0<4C_2C_6<C_4^2)$ is satisfied.
(c) Ground-state energy as a function of $\sqrt{n_{\rm c}}$. The solid lines plot Eq.\,(\ref{free_6}). The dotted lines are calculated by diagonalizing Eq.\,(\ref{H_i}).
$zt/U_0$ and $\mu/U_0$ for each line are set to the values for the same symbols in (b).}\label{phase_diagram_2}
\end{figure}%
It has been found that the SF-MI phase transition is of first order in the vicinity of the tip of the Mott lobe (see Fig.\,\ref{phase_diagram_2}) when $U_2/U_0<0.32$ \cite{kimura05}. The metastable SF and MI phases appear inside the MI and SF regions, respectively, in the phase diagram \cite{yamamoto13}. We need in this case the ground-state energy expanded up to sixth order, 
\begin{eqnarray}
E=E^0(0,n)+C_2n_{\rm c}+C_4n_{\rm c}^2+C_6n_{\rm c}^3.\label{free_6}
\end{eqnarray}
Note that $C_4$ is negative on the first-order part of the phase boundary. The right-hand side of Eq.\,(\ref{free_6}) is plotted as a function of $\sqrt{n_{\rm c}}$ in Fig.\,\ref{phase_diagram_2}\,(c).\\
\indent In the SF phase, the equation for $n_{\rm c}$,
\begin{eqnarray}
C_2n_{\rm c}+C_4n_{\rm c}^2+C_6n_{\rm c}^3=0,\label{eq:condition}
\end{eqnarray}
has a positive solution as shown in Fig.\,\ref{phase_diagram_2} (c) (see the curves with green inverted triangles and red diamonds). The ground state is thus in the SF phase if $C_4^2-4C_2C_6>0$. In the MI phase, i.e., $C_4^2-4C_2C_6<0$,  Eq.\,(\ref{eq:condition}) has a single solution $n_{\rm c}=0$ [see the curves with orange circles and blue squares in Fig.\,\ref{phase_diagram_2}(c)]. The phase boundary is, therefore, determined by the condition $C_4^2-4C_2C_6=0$.\\
\indent If $C_2>0$ in the SF phase, since the ground-state energy has a local minimum at $\sqrt{n_{\rm c}}=0$, the metastable MI state appears. 
On the other hand, the ground-state energy has a local minimum at $\sqrt{n_{\rm c}}>0$  in the MI phase, if the metastable SF state exists.
From the condition  for a local minimum $\partial E/\partial \sqrt{n_{\rm c}}=0$, we obtain \begin{eqnarray}
n_{\rm c}=0,\,\frac{-C_4\pm\sqrt{C_4^2-3C_2C_6}}{3C_6}.\label{n_}
\end{eqnarray}
The positive solution ($n_{\rm c}>0$) in Eq.\,(\ref{n_}) corresponds to the metastable SF state.
The metastable SF state thus appears if $C_4^2-3C_2C_6>0$.
Consequently, the conditions for the existence of the metastable SF and MI states are given as $3C_2C_6<C_4^2<4C_2C_6$ and $0<4C_2C_6<C_4^2$, respectively.
Figures\,\ref{phase_diagram_2}\,(a) and \ref{phase_diagram_2}\,(b) show the regions of the metastable SF and MI phases in the phase diagram when $U_2/U_0=0.31$.
The phase boundary obtained by diagonalizing Eq.\,(\ref{H_i}) agrees well with that of the perturbative mean-field theory, as shown in Fig.\,\ref{phase_diagram_2}\,(a), as far as $U_2/U_0\lesssim 0.32$.
If $0<U_2/U_0\ll 0.32$, since the SF order parameter is large even in the vicinity of the phase boundary, the perturbative expansion of the ground-state energy by the SF order parameter breaks down. The correct phase boundary is obtained by diagonalizing Eq.\,(\ref{H_i}) \cite{yamamoto13}.

\section{collective modes}\label{sec:collective_mode}
\begin{figure}%
\includegraphics[clip,width=190pt]{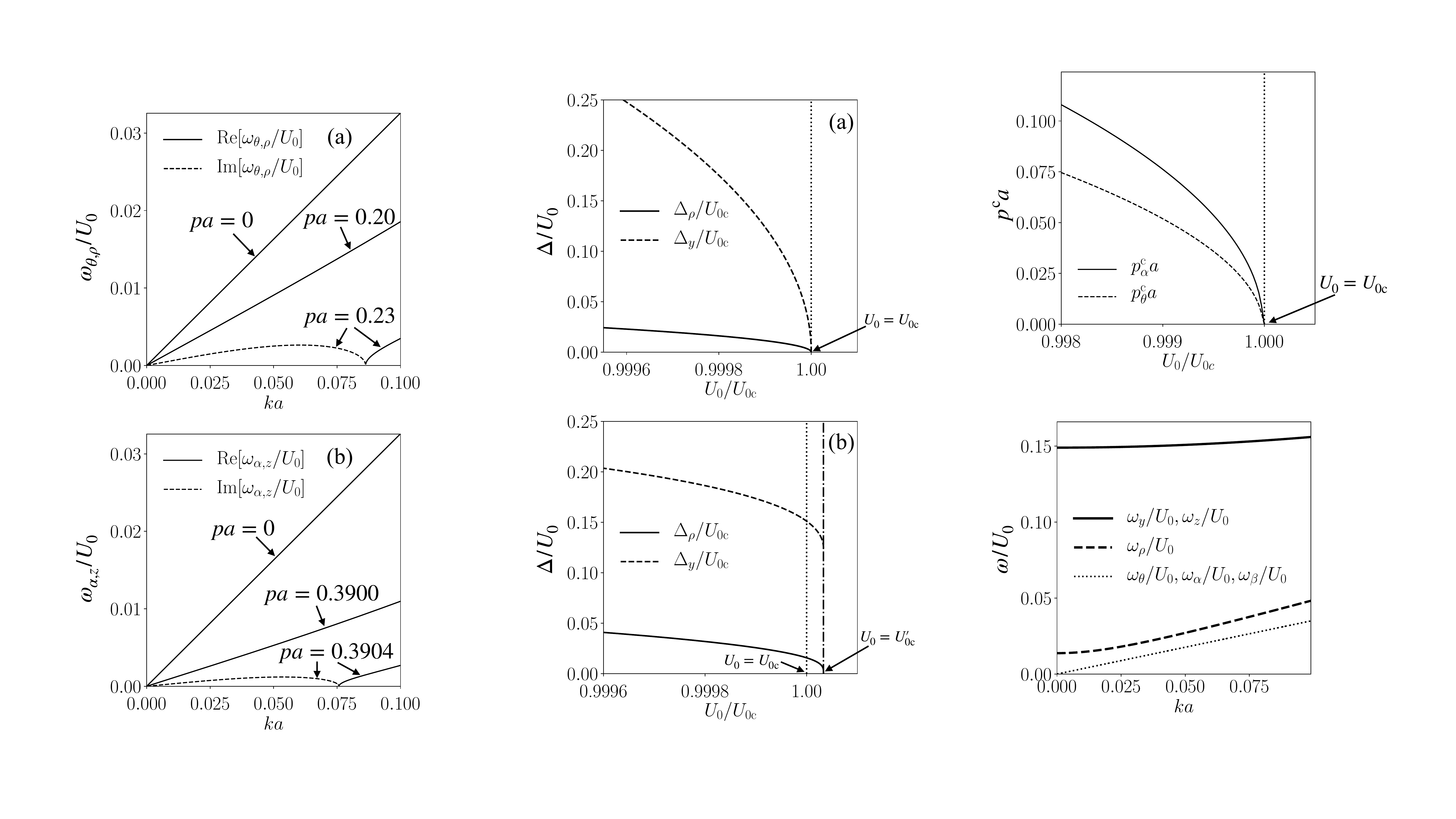}
\caption{(a) Dispersion relations for the collective modes in Eqs.\,(\ref{eq:w_theta}), (\ref{eq:w_rho}), and (\ref{eq:w_z}). We set $U_2/U_0=0.33$, $zt/U_0= 0.3834$, and $\mu/U_0=0.9831$.}  \label{fig:dispersions}
\end{figure}%
\begin{figure}%
\includegraphics[clip,width=180pt]{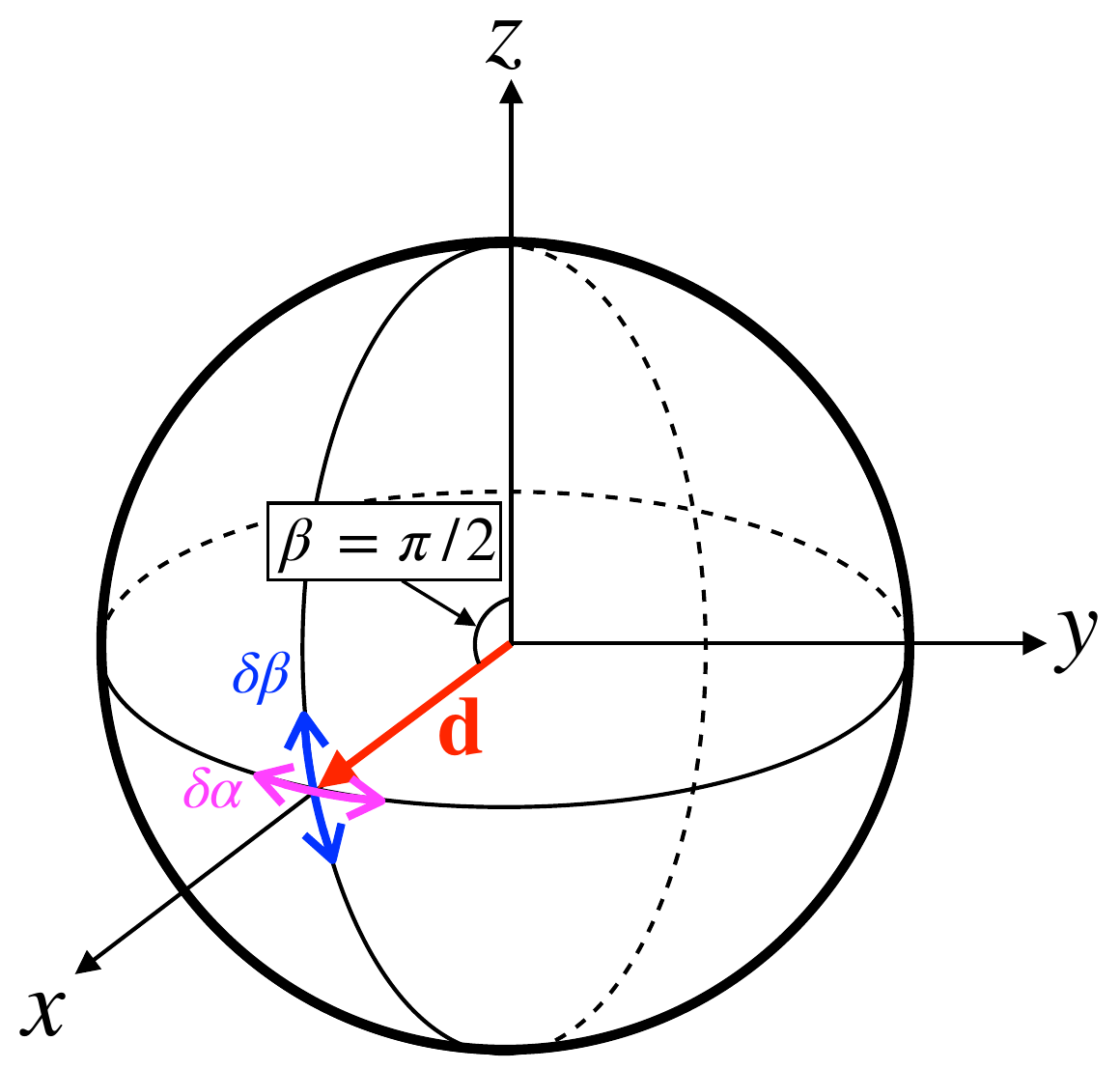}
\caption{(Color online) Schematic representation of the spin-nematic modes $\omega_\alpha$ and $\omega_\beta$ in Eqs.\,(\ref{eq:da}) and (\ref{eq:db}). $\omega_\alpha$ ($\omega_\beta$) indicates fluctuation of $\alpha$ ($\beta$).}\label{fig:d_vec}
\end{figure}%

We study collective modes in the SF phase based on the TDGL equation,
\begin{eqnarray}
iK\partial_t \Psi_\alpha
-J\partial_t^2 \Psi_\alpha
&=& 
-
\frac{\nabla^2}{2m^*}\Psi_\alpha
+
C_2 \Psi_\alpha
+
2c_4 ({\bf \Psi}^\dagger {\bf \Psi})\Psi_\alpha\nonumber \\
&&+
2c_4'\langle \! \langle \bm{F} \rangle \! \rangle \cdot \sum_\beta \bm{F}_{\alpha\beta} \Psi_\beta\nonumber \\
&&+
3c_6 ({\bf \Psi}^\dagger  {\bf \Psi})^2\Psi_\alpha +c_6'\langle \! \langle \bm{F} \rangle\! \rangle^2\Psi_\alpha \nonumber \\
&&+
2c_6'({\bf \Psi}^\dagger  {\bf \Psi})\langle \! \langle \bm{F} \rangle \! \rangle \cdot \sum_\beta \bm{F}_{\alpha\beta} \Psi_\beta,\label{TDGL}
\end{eqnarray} 
where ${\bf \Psi}=(\Psi_1,\Psi_0,\Psi_{-1})^T$ denotes the SF order parameter.
(The derivation of the TDGL equation is summarized in Appendix\,\ref{app:tdgl}). Here $\langle \! \langle \bm{F} \rangle \! \rangle = \sum_{\alpha,\beta}\Psi_{\alpha}^* \bm{F}_{\alpha\beta}\Psi_{\beta}$ is the spin average. 
\par 
Neglecting the sixth-order terms, we first examine the basic characters of collective modes within the fourth-order TDGL equation.
We assume a commensurate filling and set $K=0$ for simplicity [see Eq.\,(\ref{nui}) in Appendix \ref{sec:cons_law}]. \\
\indent We set $(\alpha,\beta,\theta)=(0,\pi/2,0)$ for the static order parameter,
\begin{eqnarray}
{\bf\Psi}^{\rm 0}
= \sqrt{\frac{\rho}{2}} 
\left[
\begin{array}{c}
-1\\
0\\
 1
 \end{array}
 \right],\label{eq:Psi_static}
 \end{eqnarray}
 where $\rho = ({{\bf \Psi}^{\rm 0}})^\dagger {\bf \Psi}^{\rm 0}=-C_2/2c_4$ is the SF density. The $d$ vector for Eq.\,(\ref{eq:Psi_static}) $\bm{d}=\sqrt{\rho}(1,0,0)^T$ is shown in Fig.\,\ref{fig:d_vec}. 
 We introduce fluctuation of the order parameter around the static solution ${\bf \Psi}^{\rm 0}$ as 
 \begin{eqnarray}
 {\bf \Psi} 
 &=& 
 \sqrt{\rho + \delta \rho} e^{i\delta \theta} R(\delta \alpha, \pi/2+\delta \beta, \delta \gamma)
 \left[
 \begin{array}{c}
 0\\
 1\\
 0
 \end{array}
 \right]\nonumber \\
 &\sim&
 {\bf \Psi}^{\rm 0} + \delta {\bf \Psi}, \label{eq:Psi+delta_Psi} 
 \end{eqnarray}
 where
 \begin{eqnarray}
 \delta {\bf \Psi} 
 =
 \left[
 \begin{array}{c}
 \delta \Psi_1\\
 \delta \Psi_0\\
 \delta \Psi_{-1}
 \end{array}
 \right]
 =
 \sqrt{\rho}
 \left[
 \begin{array}{c}
 -\frac{1}{\sqrt{2}}\left( \frac{\delta \rho}{2\rho} + i\delta \theta-i\delta \alpha \right)\\
 -\delta \beta\\
 \frac{1}{\sqrt{2}}\left( \frac{\delta \rho}{2\rho} + i \delta \theta + i\delta \alpha\right)
 \end{array}
 \right]. \nonumber \\
 \label{eq:delta_Psi}
 \end{eqnarray}
 \vspace{10pt}
The $d$ vector that corresponds to Eq.\,(\ref{eq:Psi+delta_Psi}) is given by
\begin{eqnarray} 
\bm{d}
=
\sqrt{\rho}
\left[
	\begin{array}{c}
	1\\
	0\\
	0
	\end{array} 
\right]
+
\sqrt{\rho}
\left[
	\begin{array}{c}
	\frac{\delta \rho}{2\rho} + i\delta \theta \\
	\delta \alpha \\
	-\delta \beta
	\end{array}
\right]. \label{eq:d+dd}
\end{eqnarray}
Linearizing the fourth-order TDGL equation by fluctuation $\delta {\bf \Psi}$ and Fourier transforming by $\delta \Psi_\mu=\sum_{\bm{k},\omega}  e^{i(\bm{k}\cdot \bm{r}-\omega t)}\delta \Phi_{\mu}({\bm k},\omega)\ (\mu=0,\pm1)$, we obtain
\begin{widetext}
\begin{eqnarray}
\left[
\begin{array}{cccc}
\varepsilon_k + c_+\rho & c_+\rho & c_-\rho & c_-\rho\\
c_+\rho & \varepsilon_k + c_+\rho & c_-\rho & c_-\rho \\
c_-\rho & c_-\rho &  \varepsilon_k + c_+\rho & c_+\rho \\
c_-\rho & c_-\rho & c_+ \rho &  \varepsilon_k + c_+\rho \\
\end{array}
\right]
\left[
\begin{array}{c}
\delta \Phi_1\\
\delta \Phi_1^*\\
\delta \Phi_{-1}\\
\delta \Phi_{-1}^*\\
\end{array}
\right]
&=&
J\omega^2 
\left[
\begin{array}{c}
\delta \Phi_1\\
\delta \Phi_1^*\\
\delta \Phi_{-1}\\
\delta \Phi_{-1}^*\\
\end{array}
\right], \label{eq:phi_pm}\\
\left[
\begin{array}{cc}
\varepsilon_k  +2c_4' \rho & -2c_4'\rho\\
-2c_4'\rho & \varepsilon_k + 2c_4'\rho
\end{array}
\right]
\left[
\begin{array}{c}
\delta \Phi_0\\
\delta \Phi_0^*
\end{array}
\right]
&=&
J\omega^2 
\left[
\begin{array}{c}
\delta \Phi_0\\
\delta \Phi_0^*
\end{array}
\right],\label{eq:phi_0}
\end{eqnarray}
\end{widetext}
where $\varepsilon_k\equiv \frac{\bm{k}^2}{2m^*}$ and $c_{\pm} \equiv c_4\pm c_4'$. Solving Eqs.\,(\ref{eq:phi_pm}) and (\ref{eq:phi_0}), we obtain the three degenerate gapless modes with the same dispersion,
\begin{eqnarray}
\omega_\theta (\bm k) = \omega_\alpha (\bm k) = \omega_\beta (\bm k)= \sqrt{\varepsilon_k/J}.\label{eq:w_theta} 
\end{eqnarray}
In addition, we obtain the three gapful modes that have the dispersions,
\begin{eqnarray}
\omega_\rho (\bm k)&=&\sqrt{[ \varepsilon_k + 4c_4{ \rho}]/J}\label{eq:w_rho},\\
\omega_y (\bm k) = \omega_z(\bm k) &=& \sqrt{\left[\varepsilon_k +4c_4'{ \rho} \right]/J}\label{eq:w_z}.
\end{eqnarray}
Figure \ref{fig:dispersions} shows the dispersions in Eqs.\,(\ref{eq:w_theta}), (\ref{eq:w_rho}), and (\ref{eq:w_z}). The amplitudes of the normal modes for $\omega_\theta$ and $\omega_\rho$ are given, respectively, as
\begin{eqnarray}
\delta \Phi_1-\delta \Phi_1^* -\delta \Phi_{-1}+\delta \Phi_{-1}^* &\propto& \delta \theta,\\
\delta \Phi_1+\delta \Phi_1^* -\delta \Phi_{-1}-\delta \Phi_{-1}^* &\propto& \delta \rho. 
\end{eqnarray}
They represent the phase and amplitude modes that arise due to the spontaneous breaking of ${ {\rm U}(1)}_\theta$ in Eq.\,(\ref{symmetry}) \cite{goldstone62}. \\
\indent The amplitudes of the normal modes for $\omega_{\alpha}(\bm k),\,\omega_{\beta} (\bm k)$, $\omega_y (\bm k)$, and $\omega_z(\bm k)$ are given, respectively, as 
\begin{eqnarray}
\delta \Phi_1-\delta \Phi_1^* +\delta \Phi_{-1}-\delta \Phi_{-1}^*
&\propto& \delta \langle \! \langle Q_{xy} \rangle \! \rangle
\propto \delta \alpha,\label{eq:da}\\
\delta \Phi_0 + \delta \Phi_0^* &\propto& \delta \langle \! \langle Q_{zx} \rangle \! \rangle \propto \delta \beta,\label{eq:db}\\
\delta \Phi_0 - \delta \Phi_0^*  &\propto& \delta \langle \! \langle F_y \rangle\! \rangle,\\
\delta \Phi_1+\delta \Phi_1^* +\delta \Phi_{-1}+\delta \Phi_{-1}^* &\propto& \delta \langle \! \langle F_z \rangle\! \rangle.
\end{eqnarray}
Here, we introduce the nematic tensor \cite{degennes93},
\begin{eqnarray}
Q_{\mu\nu} \equiv \frac{1}{2}( F_\mu F_{\nu} + F_{\nu} F_{\mu} ) -\delta_{\mu\nu}\frac{\bm{F}^2}{3}.
\end{eqnarray}
$Q_{\mu\nu} $ characterizes the  spin-nematic order \cite{barnett06} and $\langle \! \langle Q_{\mu\nu} \rangle \! \rangle = \sum_{\alpha,\beta}\Psi_{\alpha}^* (Q_{\mu\nu})_{\alpha\beta} \Psi_{\beta}$. $\omega_{y}$ and $\omega_{z}$ represent spin-wave modes associated with fluctuations of magnetization, while $\omega_\alpha$ and $\omega_\beta$ represent spin-nematic modes associated with fluctuations of the nematic tensor \cite{yukawa12}. They induce fluctuations of the polar and azimuthal angles of the $d$ vector $\alpha$ and $\beta$, as shown in Fig.\,\ref{fig:d_vec}. 
The spin-nematic modes $\omega_\alpha$ and $\omega_\beta$ arise due to the spontaneous breaking of ${S}^2_{\alpha,\beta}$ in Eq.\,(\ref{symmetry}).\\
\indent In the case of the first-order phase transition, dynamics of the  SF order parameter is governed by the sixth-order TDGL equation (\ref{TDGL}). In the same manner as the fourth-order TDGL equation, the dispersions of the collective modes can be derived as
\begin{eqnarray}
\omega_\rho (\bm k)&=&\sqrt{[ \varepsilon_k +4(c_4+3c_6{ \rho}'){ \rho}']/J},\label{2} \label{om_1}\\ 
\omega_y(\bm k)&=&\omega_z(\bm k)=\sqrt{[ \varepsilon_k +4(c_4' +c_6'{ \rho}'){ \rho}' ]/J}\label{3},\label{om_2}\\
\omega_\theta (\bm k)&=&\omega_{\alpha} (\bm k)=\omega_{\beta} (\bm k)=\sqrt{\varepsilon_k/J}\label{1}, \label{om_3}
\end{eqnarray}
where the superfluid density is given by
\begin{eqnarray}
{ \rho}' =  \frac{-c_4 + \sqrt{c_4^2 -3C_2c_6}}{3c_6}.
\end{eqnarray} 
\par
\begin{figure}%
\includegraphics[clip,width=210pt]{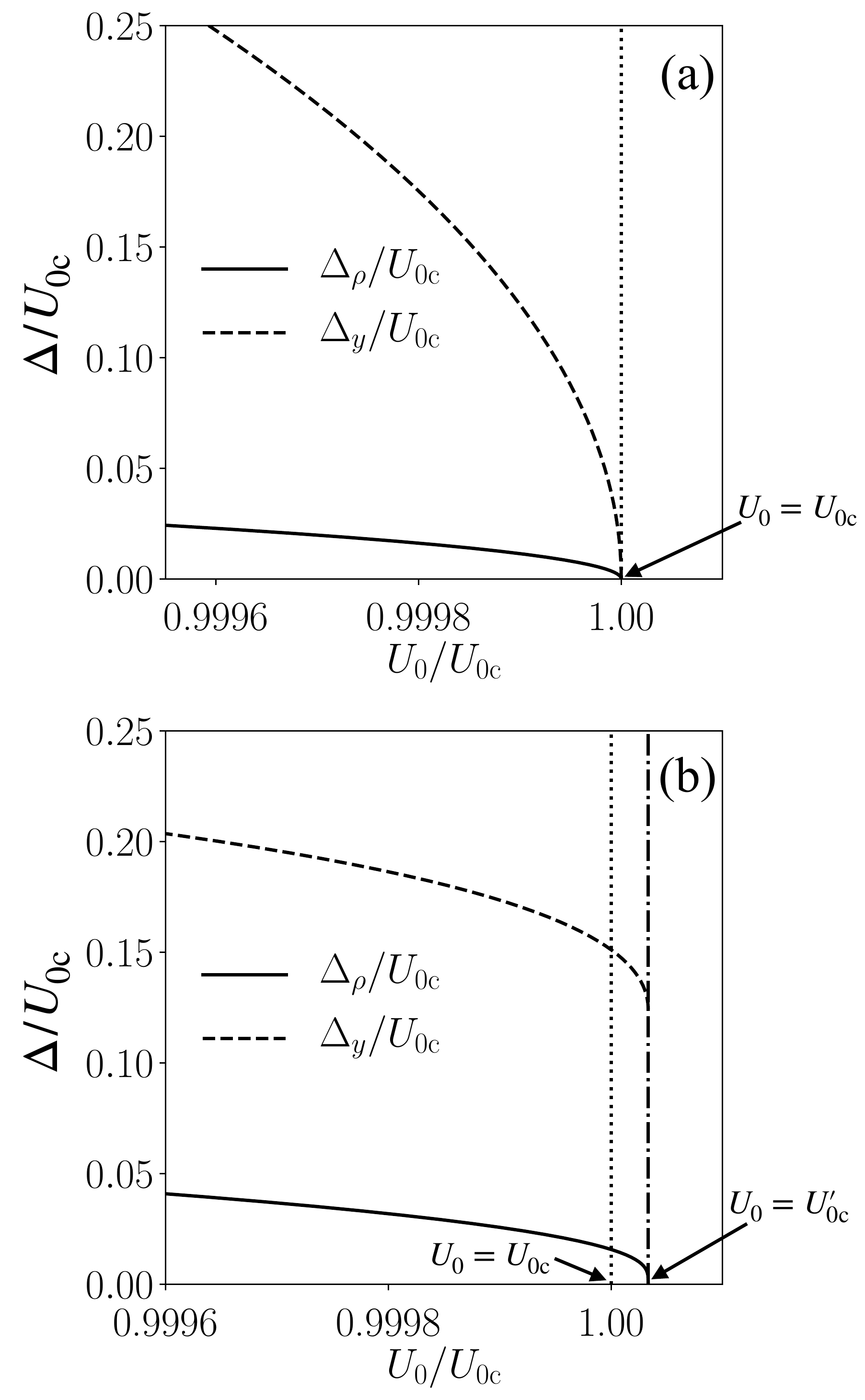}
\caption{(a) Energy gap of the gapful modes in Eqs.\,(\ref{eq:w_rho}) and (\ref{eq:w_z}) normalized by the spin-independent interaction at the second-order SF-MI phase boundary $U_{\rm 0c}$. We set $U_2/U_0=0.33$ and $\mu/U_0=0.9831$. (b) Energy gap of the gapful modes in Eqs.\,(\ref{2}) and (\ref{3}) normalized by $U_{\rm 0c}$. The metastable SF phase disappears at $U_0= U_{0{\rm c}}'$. We set $U_2/U_0=0.31$ and $\mu/U_0=1.008$.}\label{fig:gaps}
\end{figure}

Figures \,\ref{fig:gaps}\,(a) and \ref{fig:gaps}\,(b) plot the energy gap of the gapful modes $\Delta_\rho=\omega_\rho(0)$ and $\Delta_y=\omega_y(0)=\omega_z(0)$ as functions of $U_0$ near the phase boundary. $\Delta_\rho$ and $\Delta_y$ vanish at the second-order phase boundary, whereas they are finite at the first-order phase boundary. 
The finite energy gap at the phase boundary reflects the existence of the metastable SF phase. $\Delta_\rho$ vanishes at $U_0=U_{\rm 0c}'$ simultaneously with the disappearance of the metastable SF state.  
\par
The TDGL method is valid in the vicinity of the MI phase, where the order parameter is so small that the perturbative mean-field expansion is allowed. It describes low energy superfluid dynamics, in which the order parameter varies slowly in space and time.
Thus, the analytical expressions for the collective modes derived by the TDGL equation should be correct in the low energy regime for small momenta.
\par
To check the validity of the TDGL equation, we also evaluate the excitation spectrum of collective modes numerically using the time-dependent Gutzwiller ansatz \cite{navez14,shinozaki13}.
The time-dependent Gutzwiller ansatz approximates the many-body wave function as a product of single-site wave functions, each of which is expanded by finite number of Fock states.
The validity of the time-dependent Gutzwiller ansatz depends crucially on the number of Fock states involved in the expansion.
This restriction to the number of Fock states as well as the approximation for the many-body wave function as a product of single-site wave functions limits the validity of the time-dependent Gutzwiller ansatz within the low energy regime near the phase boundary, because large numbers of Fock states are required to describe the SF phase in highly excited states and/or away from the phase boundary.
The collective modes calculated by the TDGL equation and the time-dependent Gutzwiller ansatz, therefore, should agree in the low energy regime for small momenta in the vicinity of the MI phase.
\par
Figure~5 shows a comparison of the excitation spectrum calculated by the TDGL equation and the time-dependent Gutzwiller ansatz. They agree well in the low energy regime for small $k$ as expected.

\begin{figure}%
\includegraphics[clip,width=170pt]{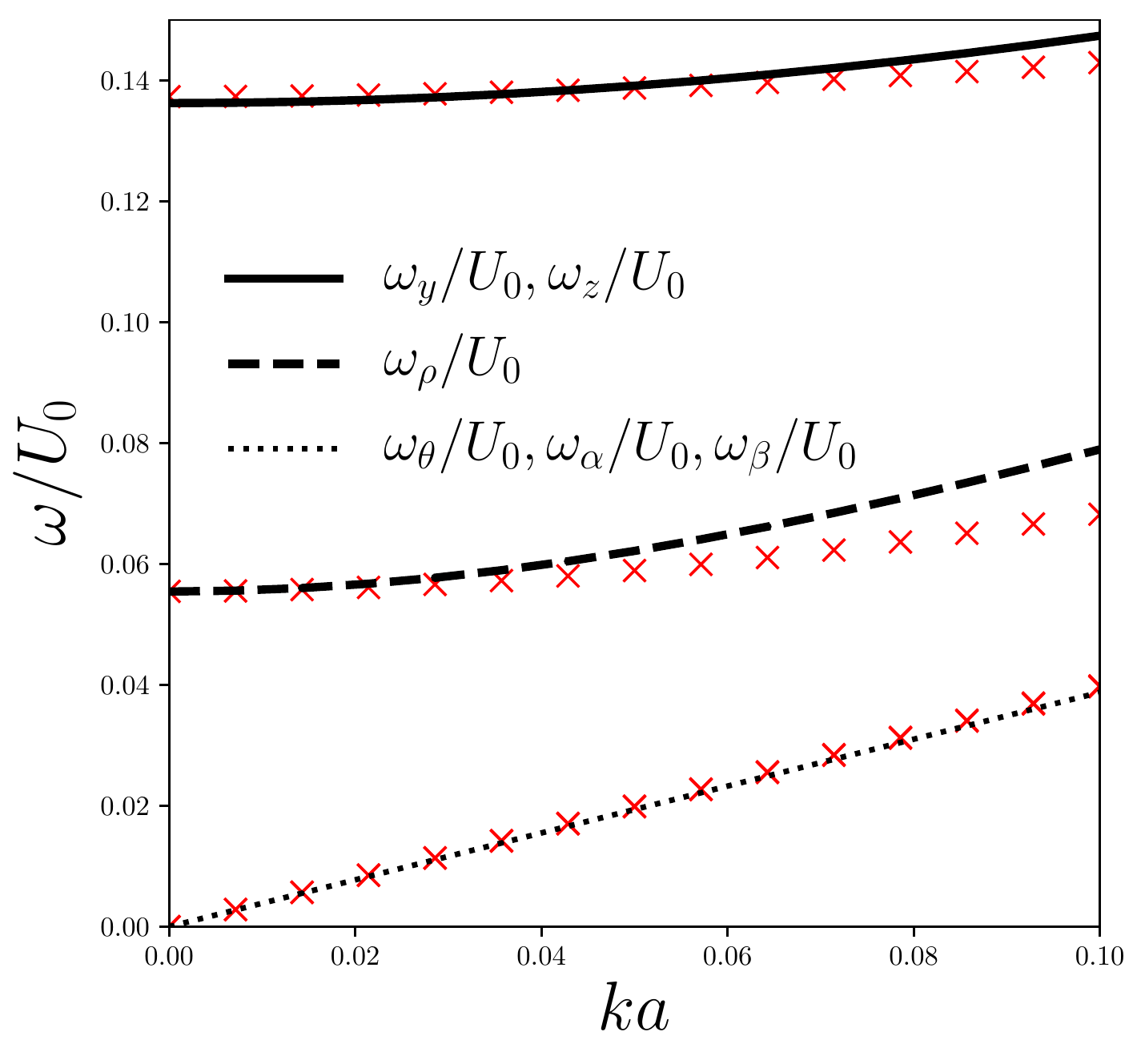}
\caption{(a) Dispersion relations for the collective modes in Eqs.\,(\ref{om_1}), (\ref{om_2}), and (\ref{om_3}). We set $U_2/U_0=0.50$, $zt/U_0= 0.451$, and $\mu/U_0=0.775$. The red crosses are the dispersion relations calculated using the time-dependent Gutzwiller ansatz. 
We expand the single-site wave function by the Fock state $|n_1,n_0,n_{-1}\rangle$ with $0\leq n_1+n_0 + n_{-1}\leq 10$. }  \label{fig:disp}
\end{figure}%
\section{Dynamical instability of supercurrents}\label{sec:mass}
\subsection{Mass currents}
To examine the stability of mass currents, we consider a current carrying state, in which $m=\pm1$ components are flowing in the same direction with momentum $\bm{p}$, as
\begin{eqnarray}
{\bf \Psi}^{\rm 0}_{\rm m}
= 
\sqrt{\frac{\bar{\rho}}{2}}e^{i\bm{p}\cdot \bm{r}}
\left[
\begin{array}{c}
-1\\
0\\
1
\end{array}
\right], \label{eq:Psi_static_mass_current}
\end{eqnarray}
 where $\bar{\rho}=-(\varepsilon_p + C_2)/2c_4$ is the SF density. Note that $p\equiv |\bm{p}|$ should be smaller than the maximum value $p_{\rm max} = \sqrt{-2m^* C_2} $ for $\bar{\rho}>0$.
 The mass and spin currents for Eq.\,(\ref{eq:Psi_static_mass_current}) are given by
\begin{eqnarray}
\bm{j}_{\rm m}&=&\bar{\rho} \frac{\bm{p}}{m^*},\\ 
\bm{j}^{\bm{n}}_{\rm s}&=&{\bf 0},
\end{eqnarray}
respectively.
Here, mass and spin currents are defined, respectively, as
\begin{eqnarray}
\bm{j}_{\rm m}&=& \frac{1}{2im^*}[{\bf \Psi}^\dagger  \nabla {\bf \Psi}-(\nabla {\bf \Psi}^\dagger) {\bf \Psi}],\label{eq;j_m}\\
\bm{j}^\mu_{\rm s} &=& \frac{1}{2im^*} [ {\bf \Psi}^\dagger F_\mu \nabla {\bf \Psi} - (\nabla {\bf \Psi}^\dagger )F_\mu {\bf \Psi}].\label{eq;j_s}
\end{eqnarray}
$\bm{j}^\mu_{\rm s}$ denotes a spin current for the $\mu$-th component of magnetization $\langle \! \langle F_\mu \rangle\! \rangle\ (\mu=x,y,z)$.   
[We derive Eqs.\,(\ref{eq;j_m}) and (\ref{eq;j_s}) in Appendix \ref{sec:cons_law}.]
We introduce fluctuation of the order parameter around the static solution ${\bf \Psi}^{\rm 0}_{\rm m}$ as
\begin{eqnarray}
{\bf \Psi}
&=& 
e^{i\bm{p}\cdot \bm{r}} \sqrt{\bar{\rho}+\delta \rho} e^{i\delta \theta} R(\delta \alpha,\pi/2+\delta \beta,\delta \gamma) 
\left[
\begin{array}{c}
0\\
1\\
0
\end{array}
\right]\nonumber  \\
&\sim& {\bf \Psi}^{\rm 0}_{\rm m} + \delta {\bf \Psi},\label{eq:Psi+delta_Psi_mass_current}
\end{eqnarray}
where
\begin{eqnarray}
\delta {\bf \Psi}
&=&
e^{i\bm{p}\cdot \bm{r}}
\left[
\begin{array}{c}
\delta \Psi_1\\
\delta \Psi_0\\
\delta \Psi_{-1}
\end{array}
\right] \nonumber \\
&=&
\sqrt{\bar{\rho}}
e^{i\bm{p}\cdot \bm{r}}
\left[
\begin{array}{c}
-\frac{1}{\sqrt{2}} \left(\frac{\delta \rho}{2\bar{\rho}}+ i\delta \theta - i\delta \alpha\right)\\
-\delta \beta\\
\frac{1}{\sqrt{2}} \left(\frac{\delta \rho}{2\bar{\rho}}+ i\delta \theta + i\delta \alpha\right)
\end{array}
\right].
\end{eqnarray}
Substituting Eq.\,(\ref{eq:Psi+delta_Psi_mass_current}) into the fourth-order TDGL equation with $K=0$ and linearizing with respect to $\delta {\bf \Psi}$, we obtain
\begin{widetext}
\begin{eqnarray}
\left[
\begin{array}{cccc}
\varepsilon_k + \frac{\bm{p}\cdot\bm{k}}{m^*}+ c_+\bar{\rho} & c_+\bar{\rho} & c_-\bar{\rho} & c_-\bar{\rho}\\
c_+\bar{\rho} & \varepsilon_k -\frac{\bm{p}\cdot\bm{k}}{m^*}+ c_+\bar{\rho} & c_-\bar{\rho} & c_-\bar{\rho} \\
c_-\bar{\rho} & c_-\bar{\rho} &  \varepsilon_k +\frac{\bm{p}\cdot\bm{k}}{m^*}+ c_+\bar{\rho} & c_+\bar{\rho} \\
c_-\bar{\rho} & c_-\bar{\rho} & c_+\bar \rho &  \varepsilon_k -\frac{\bm{p}\cdot\bm{k}}{m^*}+ c_+\bar{\rho} \\
\end{array}
\right]
\left[
\begin{array}{c}
\delta {\Phi}_1\\
\delta {\Phi}_1^*\\
\delta {\Phi}_{-1}\\
\delta {\Phi}_{-1}^*\\
\end{array}
\right]
&=&
J\omega^2 
\left[
\begin{array}{c}
\delta {\Phi}_1\\
\delta {\Phi}_1^*\\
\delta {\Phi}_{-1}\\
\delta {\Phi}_{-1}^*\\
\end{array}
\right],
\label{eq:phi_pm_mass}
\end{eqnarray}

\begin{eqnarray}
\left[
\begin{array}{cc}
\varepsilon_k  +\frac{\bm{p}\cdot \bm{k}}{m^*}+2c_4' \bar{\rho} & -2c_4'\bar{\rho}\\
-2c_4'\bar{\rho} & \varepsilon_k -\frac{\bm{p}\cdot \bm{k}}{m^*}+ 2c_4'\bar{\rho}
\end{array}
\right]
\left[
\begin{array}{c}
\delta {\Phi}_0\\
\delta {\Phi}_0^*
\end{array}
\right]
&=&
J\omega^2 
\left[
\begin{array}{c}
\delta {\Phi}_0\\
\delta {\Phi}_0^*
\end{array}
\right].
\label{eq:phi_0_mass}
\end{eqnarray}
\end{widetext}
From Eqs.\,(\ref{eq:phi_pm_mass}) and (\ref{eq:phi_0_mass}), we obtain the three gapless modes with the dispersions,
\begin{eqnarray}
\omega_{\theta,\rho} (\bm k, \bm p)
&=
& \sqrt{ \left[\varepsilon_k + 2c_4 \bar{\rho} - \sqrt{(2c_4 \bar \rho)^2 + \left(\frac{\bm p\cdot \bm k}{m^*}\right)^2} \right]/J}, \nonumber \\\label{eq:omega_theta,rho}\\
\omega_{\alpha,z} (\bm k, \bm p)
&=&
\omega_{\beta,y}(\bm{k},\bm{p})\nonumber \\
&=&
 \sqrt{ \left[\varepsilon_k + 2c_4' \bar{\rho} - \sqrt{(2c_4' \bar \rho)^2 + \left(\frac{\bm p\cdot \bm k}{m^*}\right)^2} \right]/J}.\nonumber \\
\label{eq:omega_alpha,z}
\end{eqnarray}
The amplitudes of the normal modes for $\omega_{\theta,\rho}$, $\omega_{\alpha,z}$, and $\omega_{\beta,y}$ are given, respectively, as
\begin{eqnarray} 
&&L_{\theta,\rho} \frac{\delta \rho}{\bar \rho} + L_{\theta,\theta} \delta \theta,\\
&& L_{\alpha,z} \frac{\delta \langle \! \langle F_z \rangle \! \rangle }{\bar \rho} + L_{\alpha,\alpha} \frac{\delta \langle \! \langle Q_{xy} \rangle \! \rangle }{\bar{\rho}},\\ 
&& L_{\beta,y} \frac{\delta \langle \! \langle F_y \rangle \! \rangle }{\bar \rho} + L_{\beta,\beta}  \frac{\delta \langle \! \langle Q_{xy} \rangle \! \rangle }{\bar{\rho}},
\end{eqnarray}
where
\begin{eqnarray}
L_{\theta,\theta}
&=&
\frac{1}{2}
 \left[
 1- \frac{\bm{p}\cdot \bm{k}}{2c_4 \bar{\rho} m^*} 
 +\sqrt{1+\left( \frac{\bm{p}\cdot \bm{k}}{2c_4 \bar{\rho} m^*}\right)^2}
 \right],\nonumber \\ \label{eq:c_const_1}\\
L_{\theta,\rho}
&=&
\frac{i}{4}\left[1+ \frac{\bm{p}\cdot \bm{k}}{2c_4 \bar{\rho} m^*} -\sqrt{1+\left( \frac{\bm{p}\cdot \bm{k}}{2c_4 \bar{\rho} m^*}\right)^2}\right],\nonumber \\ \\
L_{\alpha,\alpha}&=&L_{\beta,\beta} \nonumber \\
&=&
\frac{1}{2}
 \left[
 1- \frac{\bm{p}\cdot \bm{k}}{2c'_4 \bar{\rho} m^*} +\sqrt{1+\left( \frac{\bm{p}\cdot \bm{k}}{2c'_4 \bar{\rho} m^*}\right)^2}
 \right],\nonumber \\ \\
L_{\alpha,z}&=&L_{\beta,y} \nonumber \\
&=&
\frac{i}{4}
 \left[
 1+ \frac{\bm{p}\cdot \bm{k}}{2c'_4 \bar{\rho} m^*} -\sqrt{1+\left( \frac{\bm{p}\cdot \bm{k}}{2c'_4 \bar{\rho} m^*}\right)^2}
 \right].\nonumber \\ \label{eq:c_const_2}
\end{eqnarray}
\begin{figure}
\includegraphics[clip,width=200pt]{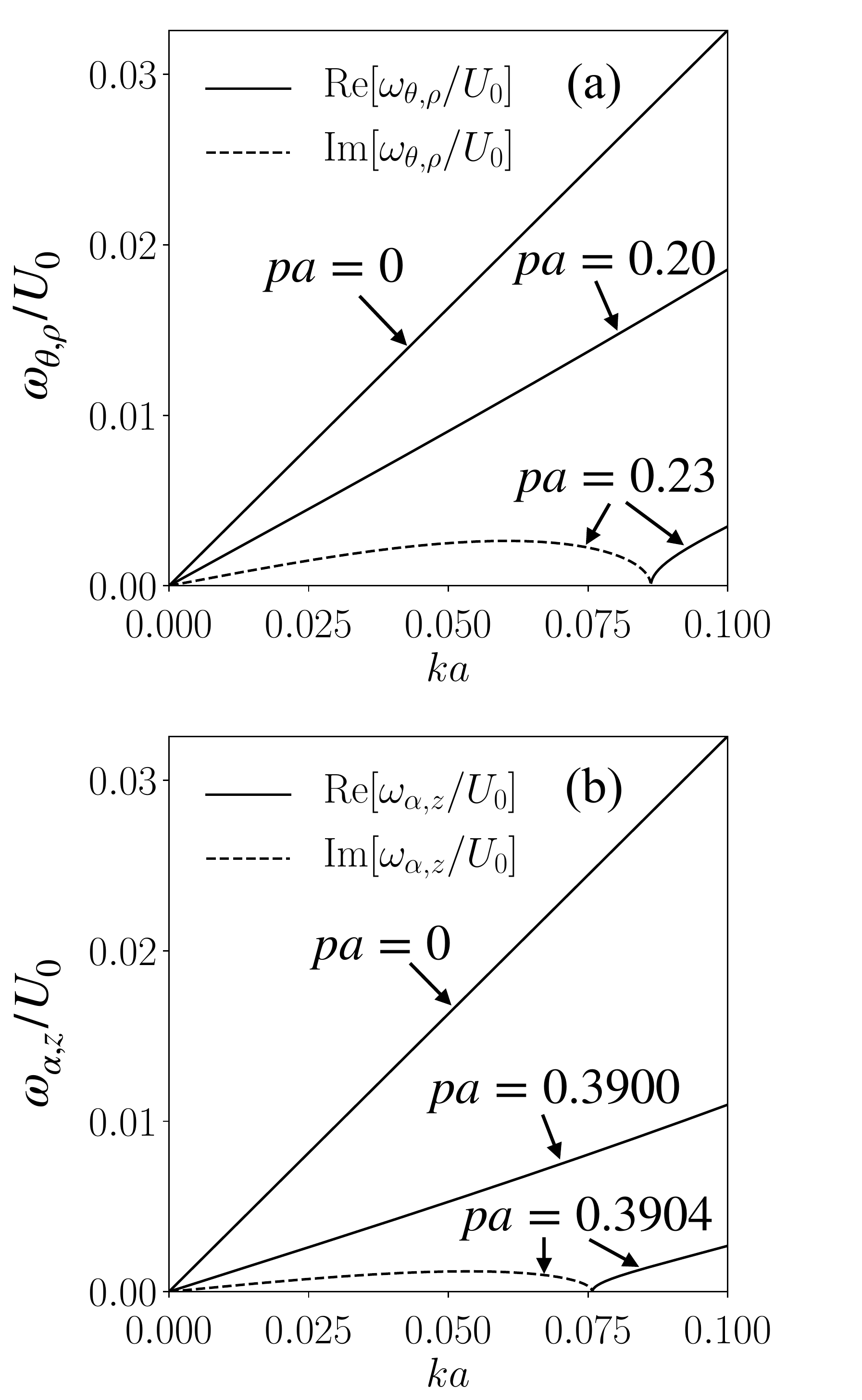}
\caption{Dispersion relations for the collective modes in Eqs.\,(\ref{eq:omega_theta,rho}) and (\ref{eq:omega_alpha,z}) in the presence of a mass current with momentum $\bm{p}$ . We set $U_2/U_0=0.33$, $zt/U_0=0.3834$, and $\mu/U_0=0.9831$. The solid lines in (a) and (b) denote the real parts of Eqs.\,(\ref{eq:omega_theta,rho}) and (\ref{eq:omega_alpha,z}), respectively. The dashed lines in (a) and (b) indicate the imaginary parts of Eqs.\,(\ref{eq:omega_theta,rho}) and (\ref{eq:omega_alpha,z}), respectively.}
\label{dispersion2}
\end{figure}

\begin{figure}%
\includegraphics[clip,width=210pt]{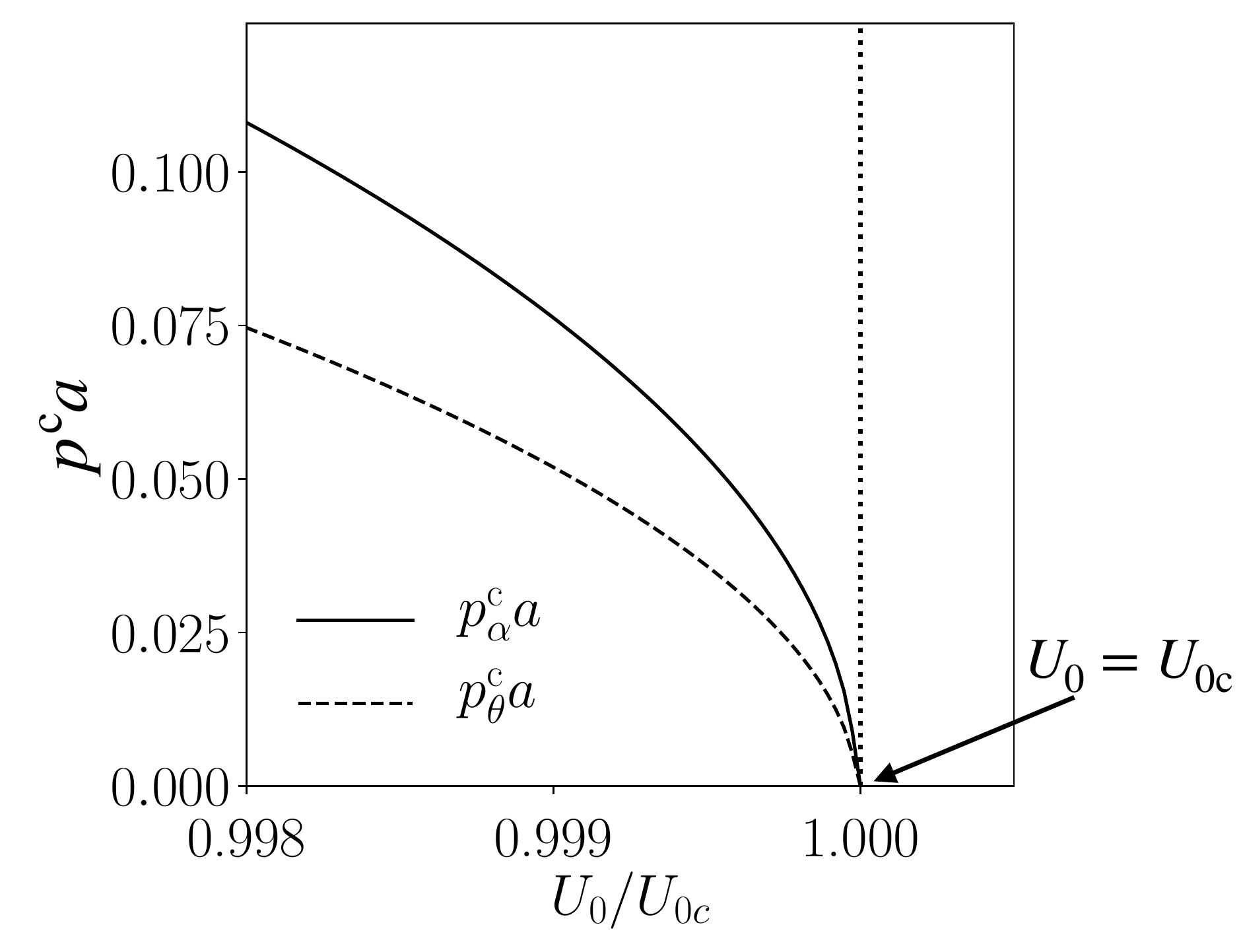}
\caption{Critical momenta for mass currents $p_\theta^{\rm c}$ and $p_\alpha^{\rm c}$ in Eqs.\,(\ref{pc1}) and (\ref{pc1'}) as functions of $U_0/U_{\rm 0c}$ for the second-order phase transition. $U_{\rm 0c}$ is the critical value of the spin-independent interaction for the SF-MI phase transition. We set $U_2/U_0=0.33$ and $\mu/U_0=0.9831$.}\label{fig:pc}
\end{figure}
\indent Since the phase factor $e^{i\bm p \cdot \bm r}$ in Eq.\,(\ref{eq:Psi_static_mass_current}) breaks the particle-hole symmetry, a pair of gapful mode and a gapless mode in Eqs.\,(\ref{eq:w_theta}), (\ref{eq:w_rho}), and (\ref{eq:w_z}) are coupled in the presence of a mass current and yields a single gapless mode. $\omega_\rho$ and $\omega_\theta$ are coupled, for example, and yield $\omega_{\theta,\rho}$. \\
\indent Figure \ref{dispersion2} shows the evolution of the dispersions in Eqs.\,(\ref{eq:omega_theta,rho}) and (\ref{eq:omega_alpha,z}), as $p$ is increased.
Here, we set $\bm{k}$ being antiparallel to $\bm{p}$. 
If $p$ exceeds the critical momenta for dynamical instabilities, the dispersion of collective modes acquires an imaginary part.
The critical momenta $p_{\theta}^{\rm c}$ and $p_{\alpha}^{\rm c}$ for the onset of the dynamical instabilities induced by $\omega_{\theta,\rho}$ and $\omega_{\alpha,z}$ are given, respectively, by  
\begin{eqnarray}
&&p_{\theta}^{\rm c}=\sqrt{-\frac{2}{3}C_2m^*}= \frac{p_{\rm max}}{ \sqrt{3}}\label{pc1},\\
&&p_{\alpha}^{\rm c} = \sqrt{-\frac{2C_2c_4'm^*}{2c_4'+c_4}}= p_{\rm max} \sqrt{\frac{c_4'}{2c_4'+c_4}}.\label{pc1'}
\end{eqnarray}
The critical momentum for $\omega_{\beta,y}$ is equivalent to $p_\alpha^{\rm c}$.
Figure \ref{fig:pc} shows the critical momenta $p_\theta^{\rm c}$ and $p_{\alpha}^{\rm c}$ as functions of $U_0$ in the vicinity of the second-order phase transition.
The dynamical instability induced by $\omega_{\theta,\rho}$ for $p>p_\theta^{\rm c}$ leads to exponential growth of $\delta \theta$ and $\delta \rho$ in time.
\begin{figure}%
\includegraphics[clip,width=210pt]{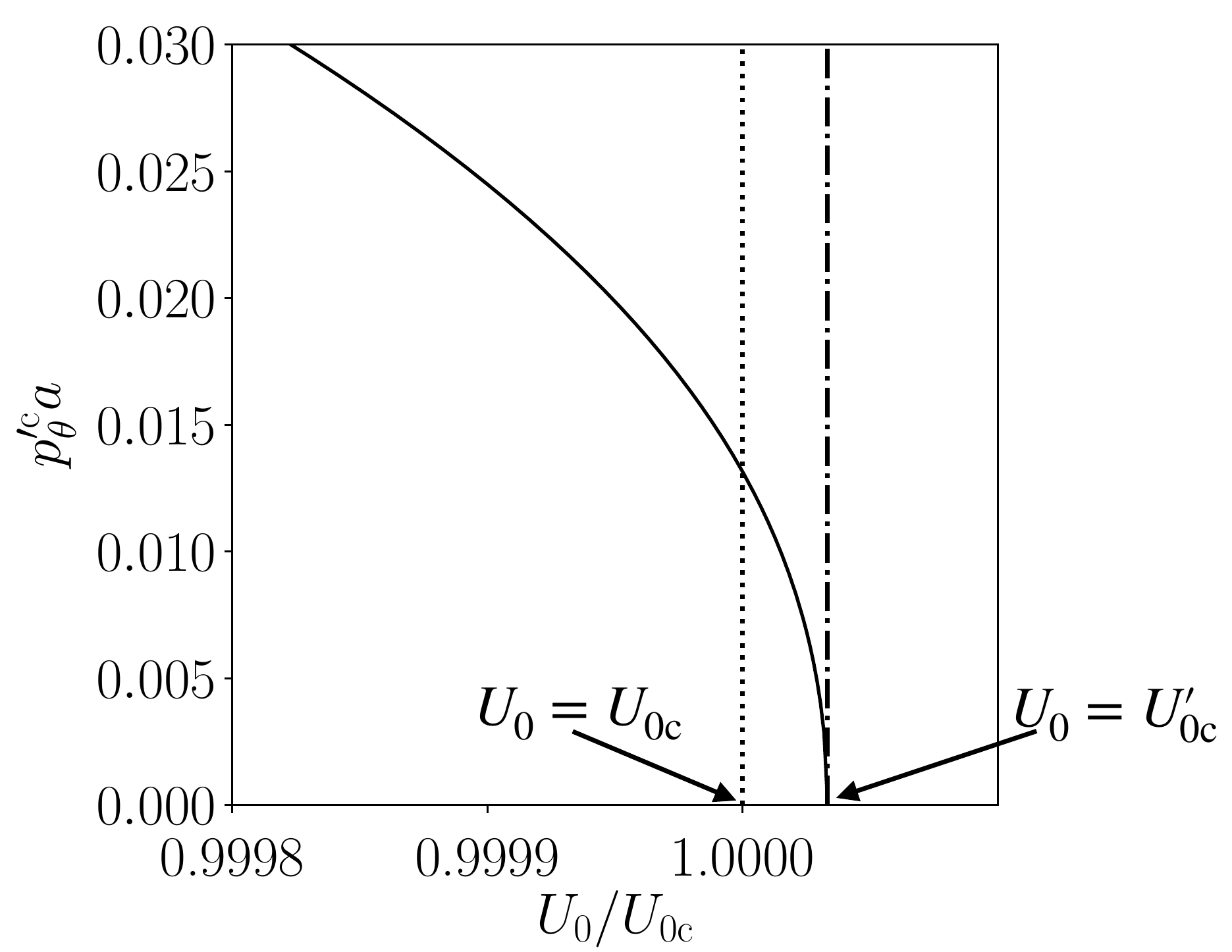}
\caption{Critical momentum for mass currents ${p_\theta'}^{\rm c}$ in Eq.\,(\ref{eq:pc_1st}) as a function of $U_0/U_{\rm 0c}$ for the first-order phase transition. $U_{\rm 0c}$ is the critical value of the spin-independent interaction for the SF-MI phase transition. $U_{\rm 0c}'$ is the value of the spin-independent interaction at the boundary, at which the metastable SF phase disappears. We set $U_2/U_0=0.31$ and $\mu/U_0=1.008$. }
\label{net_current}
\end{figure}
Meanwhile, the dynamical instabilities induced by $\omega_{\alpha,z}$ and $\omega_{\beta,y}$ for $p>p_\alpha^{\rm c}$ lead to exponential growth of $\delta \langle \! \langle F_y \rangle\! \rangle$, $\delta \langle \! \langle F_z \rangle\! \rangle$, $\delta \langle \! \langle Q_{xy} \rangle\! \rangle$, and $\delta \langle \! \langle Q_{zx} \rangle\! \rangle$ in time. \\
\indent In the case of the first-order phase transition, using the sixth-order TDGL equation (\ref{TDGL}), the dispersions for the gapless modes are obtained as
\begin{eqnarray}
\omega_{\theta,\rho}(\bm{k},\bm{p})
&=&
 \left[
 	\varepsilon_k
	+ 2 (c_4 + 3c_6{\bar \rho}'){\bar \rho}' 
	\right.\nonumber \\
	&&\left.
	-\sqrt{
		(\bm k \cdot \bm p/m^*)^2 + 4 (c_4 + 3c_6{\bar \rho}')^2{\bar \rho}^{'2}
		}
	\right]^{1/2},\nonumber \\ \label{eq:omega'_tp}\\
\omega_{\alpha,z}(\bm{k},\bm{p})
&=&
\omega_{\beta,y}(\bm{k},\bm{p})\nonumber \\
&=&
 \left[
 \varepsilon_k 
+ 2 (c_4' + c_6'{\bar \rho}'){\bar \rho}' 
\right.\nonumber \\
&&
\left.
-\sqrt{
(\bm k \cdot \bm p/m^*)^2 + 4 (c_4' + c_6'{\bar \rho}')^2{\bar \rho}^{'2}}
\right]^{1/2},\nonumber \\ \label{eq:omega'_az}
\end{eqnarray}
where the superfluid density is given by
\begin{eqnarray}
\bar{\rho}' = \frac{-c_4 + \sqrt{ c_4^2 -3c_6(C_2 + \varepsilon_p)}}{3c_6}.
\end{eqnarray}
The maximum value of the momentum $p_{\rm max}'$, which satisfies $\bar{\rho}'>0$, is given by
\begin{eqnarray}
p_{\rm max}' = \sqrt{2m^* \left(  \frac{c_4^2}{3c_6} -C_2\right) }.
\end{eqnarray} 
From Eq.\,(\ref{eq:omega'_tp}), the critical momentum ${p_\theta'}^{\rm c}$ for the onset of dynamical instabilities induced by $\omega_{\theta, \rho}$ is 
\begin{eqnarray}
{p_\theta'}^{\rm c}
=
 \sqrt{\frac{m^*}{4}
 \left[ 
 	\frac{c_4^2}{c_6} - 4C_2 
	+ \sqrt{
		\frac{c_4^2}{c_6} 
		\left(
			\frac{c_4^2}{c_6} -\frac{8}{3}C_2 
		\right)
	}
\right]
}.\label{eq:pc_1st}
\end{eqnarray}
Figure\,\ref{net_current} shows the critical momenta $p^{\rm c}_\theta$ and ${p_\theta'}^{\rm c}$  as functions of $U_0$.\\
\indent In the case of the second-order phase transition, we find $p_\theta^{\rm c}\rightarrow 0$ as $U_0\rightarrow U_{\rm 0c}$, as shown in Fig.\,\ref{fig:pc}. In fact, from Eq.\,(\ref{pc1}), one finds $p_\theta^{\rm c}=0$ due to $C_2=0$ at the phase boundary.
This can be understood as follows:
The dynamical instability is expected to occur if the phase gradient per healing length $\xi$ exceeds $\pi/2$, i.e., $p\xi>\pi/2$, due to negative effective mass \cite{altman05}.
Since the healing length $\xi=1/\sqrt{4|C_2|m^*}$ diverges as $|C_2|\rightarrow 0$ at the phase boundary, $p_\theta^{\rm c}$ also vanishes at $U_0=U_{\rm 0c}$ \cite{healing_length}.
Meanwhile, in the case of the first-order phase transition, Fig.\,\ref{net_current} shows that ${p_\theta'}^{\rm c}>0$ at the phase boundary reflecting the finite healing length $\xi=1/\sqrt{8m^*(|c_4|\rho'-C_2)}$ \cite{healing_length}. Furthermore, it is finite throughout the metastable SF phase and vanishes at $U_0=U_{\rm 0c}'$, at which $C_4^2=3C_2C_6$ holds from Eq.\,(\ref{n_}) and the healing length $\xi$ diverges. The finite critical momentum means that the metastable SF state can sustain a mass current without dissipation as far as $p<{p_\theta'}^{\rm c}$ and therefore exhibits superfluidity.\\
\indent In contrast to the case of the second-order phase transition, the dispersions of $\omega_{\alpha,z}$ and $\omega_{\beta,y}$ exhibit very little change in the presence of a finite mass current. Thus, in the case of the first-order phase transition, $\omega_{\alpha,z}$ and $\omega_{\beta,y}$ do not cause dynamical instabilities. 
\subsection{Spin currents}\label{sec:spin_current}
To study the stability of spin currents, we assume a static solution, in which $m=\pm1$ components flowing in the opposite direction with the same momenta $\bm{p}$, as
\begin{eqnarray}
{\bf \Psi}^{\rm 0}_{\rm s} 
=
\sqrt{\frac{\bar{\rho}}{2}}
\left[
\begin{array}{c}
-e^{-i\bm{p}\cdot \bm{r}}\\
0\\
e^{i\bm{p}\cdot \bm{r}}
\end{array}
\right], \label{eq:Psi_static_js}
\end{eqnarray}
where $\bar{\rho}=-(\varepsilon_p+C_2)/2c_4$ is the SF density.
Since $m=\pm1$ components flow in the opposite direction with the same momenta, the net mass current vanishes ($\bm{j}_{\rm m}={\bf 0}$). The spin current for Eq.\,(\ref{eq:Psi_static_js}) is given as
\begin{eqnarray}
\bm{j}^\mu_{\rm s} = -\frac{\bm{p}}{m^*}\delta_{\mu,z}.\label{eq:j_s_}
\end{eqnarray}\\
\indent We introduce fluctuation of the order parameter around the static solution as
\begin{eqnarray}
{\bf \Psi} = {\bf \Psi}^{\rm 0}_{\rm s} + \delta {\bf \Psi}, \label{eq:Psi+delta_Psi_js}
\end{eqnarray}
where
\begin{eqnarray}
\delta {\bf \Psi} 
&=&
\left[
\begin{array}{c}
\delta \Psi_1 e^{-i\bm{p}\cdot \bm{r}}\\
\delta \Psi_0\\
\delta \Psi_{-1} e^{i\bm{p}\cdot \bm{r}}
\end{array}
\right]\nonumber 
\end{eqnarray}
\begin{eqnarray}
=\sqrt{\bar{\rho}}
\left[
\begin{array}{c}
-\frac{e^{-i\bm{p}\cdot \bm{r}}}{\sqrt{2}} \left( \frac{\delta \rho}{2\bar{\rho}} + i\delta \theta - i\delta \alpha \right)\\
-\delta \beta\\
\frac{e^{i\bm{p}\cdot \bm{r}}}{\sqrt{2}} \left( \frac{\delta \rho}{2\bar{\rho}} + i\delta \theta + i\delta \alpha \right)
\end{array}
\right].
\end{eqnarray}\\
\indent Substituting Eq.\,(\ref{eq:Psi+delta_Psi_js}) into the fourth-order TDGL equation with $K=0$ and linearizing with respect to fluctuation $\delta {\bf \Psi}$, we obtain the equations for the Fourier components as
\begin{widetext}
\begin{eqnarray}
\left[
\begin{array}{cccc}
\varepsilon_k + \frac{\bm{p}\cdot\bm{k}}{m^*}+ c_+\bar{\rho} & c_+\bar{\rho} & c_-\bar{\rho} & c_-\bar{\rho}\\
c_+\bar{\rho} & \varepsilon_k -\frac{\bm{p}\cdot\bm{k}}{m^*}+ c_+\bar{\rho} & c_-\bar{\rho} & c_-\bar{\rho} \\
c_-\bar{\rho} & c_-\bar{\rho} &  \varepsilon_k -\frac{\bm{p}\cdot\bm{k}}{m^*}+ c_+\bar{\rho} & c_+\bar{\rho} \\
c_-\bar{\rho} & c_-\bar{\rho} & c_+ \bar \rho&  \varepsilon_k +\frac{\bm{p}\cdot\bm{k}}{m^*}+ c_+\bar{\rho} \\
\end{array}
\right]
\left[
\begin{array}{c}
\delta {\Phi}_1\\
\delta {\Phi}_1^*\\
\delta {\Phi}_{-1}\\
\delta {\Phi}_{-1}^*\\
\end{array}
\right]
&=&
J\omega^2 
\left[
\begin{array}{c}
\delta {\Phi}_1\\
\delta {\Phi}_1^*\\
\delta {\Phi}_{-1}\\
\delta {\Phi}_{-1}^*\\
\end{array}
\right],\label{eq:phi_pm_spin}
\end{eqnarray}
\end{widetext}
\begin{eqnarray}
\left[
\begin{array}{cc}
\varepsilon_k  -\varepsilon_p +2c_4' \bar{\rho} & -2c_4'\bar{\rho}\\
-2c_4'\bar{\rho} & \varepsilon_k -\varepsilon_p+ 2c_4'\bar{\rho}
\end{array}
\right]
\left[
\begin{array}{c}
\delta {\Phi}_0\\
\delta {\Phi}_0^*
\end{array}
\right]
&=&
J\omega^2 
\left[
\begin{array}{c}
\delta {\Phi}_0\\
\delta {\Phi}_0^*
\end{array}
\right].\nonumber \\ \label{eq:phi_0_spin}
\end{eqnarray}
From Eq.\,(\ref{eq:phi_0_spin}), we find that $\omega_\beta$ takes the form,
\begin{eqnarray}
{\omega}_{\beta} (k,p) = \sqrt{ ( \varepsilon_k - \varepsilon_p)/J}. \label{eq:tilde_omega_beta}
\end{eqnarray}
The amplitude of the normal mode for $\omega_\beta$ is
\begin{eqnarray}
\delta {\Phi}_0 + \delta {\Phi}_0^* \propto \delta \beta \ (\propto \delta \langle \! \langle Q_{zx} \rangle \! \rangle ). \label{eq:delta_beta_}
\end{eqnarray}
Equation (\ref{eq:tilde_omega_beta}) shows that ${\omega}_{\beta}(k<p,p)$ has an imaginary part for any $p\neq 0$.
This indicates that even an infinitesimally small amount of spin current induces dynamical instabilities and leads to exponential growth of the long wave length spin-nematic fluctuation.
We also find that $\omega_\beta$ derived from the sixth-order TDGL equation (\ref{TDGL}) with $K=0$ has the same dispersion relation as Eq.\,(\ref{eq:tilde_omega_beta}). The critical momentum is thus also zero. The zero critical velocity for spin currents has been also reported in the study of a spin-1 BEC \cite{fujimoto12,zhu15}.
\section{Stability of spin currents}
\begin{figure}
\includegraphics[clip,width=250pt]{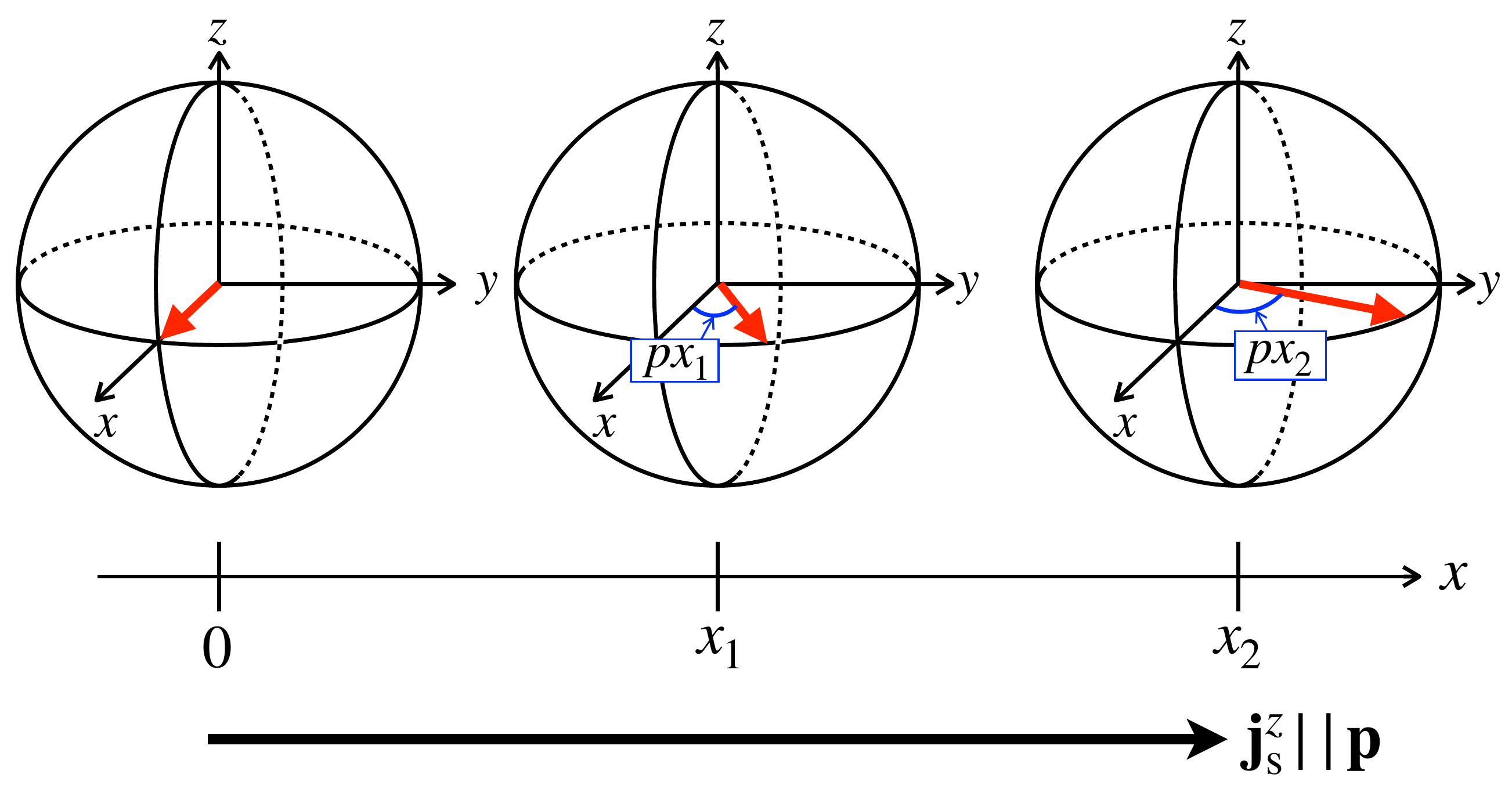}%
\caption{(Color online) Schematic illustration of the spatial variation of the $d$ vector in Eq.\,(\ref{d_}) and the spin current $\bm{j}_{\rm s}^z$ in Eq.\,(\ref{jsz}). We set $\beta=\pi/2$ and $\bm{p}=(p,0,0)^T$.}\label{fig:d_and_js}
\end{figure}
 \begin{figure}
\includegraphics[clip,width=250pt]{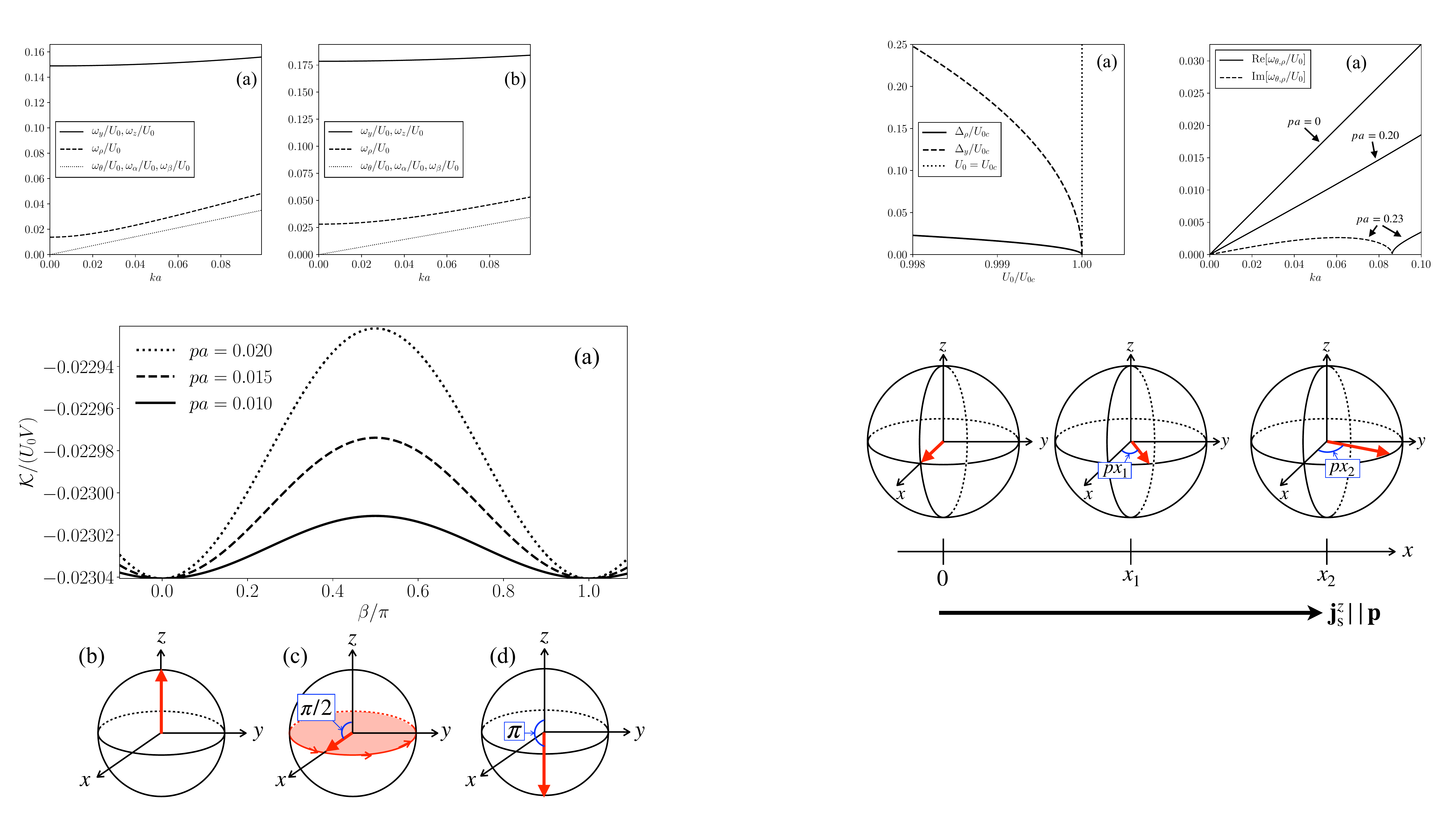}%
\caption{(Color online) (a) Energy functional (\ref{eq:energy_functional}) as a function of $\beta$. We set $U_2/U_0=0.33,\ zt/U_0=0.3834$, and $\mu/U_0=0.9831$. (b), (c), and (d) are the schematic drawings of $d$ vectors (\ref{d_}) for $\beta=0,\pi/2,$ and $\pi$, respectively.}\label{fig:energy}
\end{figure}
To understand the origin of the dynamical instability of spin currents, we consider the static solution,
\begin{eqnarray}
{\bf \Psi}_{\bm{p},\beta}
=
\sqrt{\rho}
\left[
\begin{array}{c}
-\frac{e^{-i\bm{p} \cdot \bm{r}}}{\sqrt{2}} \sin \beta\\
\cos \beta\\
\frac{e^{i\bm{p} \cdot \bm{r}}}{\sqrt{2}} \sin \beta
\end{array}
\right],\label{eq:Psi_js}
\end{eqnarray}
where ${\bf \Psi}_{\bm{p},\beta=\pi/2}={\bf \Psi}_{\rm s}^{\rm 0}$.
The $d$ vector for Eq.\,(\ref{eq:Psi_js}) is given by
\begin{eqnarray}
\bm{d} 
= 
R_z(\bm{p}\cdot \bm{r})
\sqrt{\rho}
\left[
\begin{array}{c}
\sin \beta\\
0\\
\cos \beta
\end{array}
\right],\label{d_}
\end{eqnarray}
where $R_z(\theta)$ represents the rotation matrix by angle $\theta$ about the $z$-axis.\\
\indent The SF density $\rho$ and spin currents $\bm{j}^\mu_{\rm s}$ for Eq.\,(\ref{eq:Psi_js}) are given as
\begin{eqnarray}
\rho &=& -\frac{C_2 + \varepsilon_p \sin^2 \beta}{2c_4},\\
\bm{j}_{\rm s}^x&=& \frac{\bm{p}}{m^*} \rho \sin\beta \cos\beta \cos(\bm{p}\cdot\bm{r}),\label{jsx}\\
\bm{j}_{\rm s}^y&=& \frac{\bm{p}}{m^*} \rho \sin\beta \cos\beta \sin(\bm{p}\cdot\bm{r}),\label{jsy}\\
\bm{j}_{\rm s}^z&=& -\frac{\bm{p}}{m^*}\sin^2 \beta.\label{jsz}
\end{eqnarray}
Equations (\ref{jsx})-(\ref{jsz}) demonstrate that spatial variation of the $d$ vector induces a spin current. When $\beta=\pi/2$, for example, as one moves in the direction of $\bm{p}$, the $d$ vector rotating  in the $xy$ plane induces a spin current $\bm{j}^z_{\rm s}||\bm{p}$, as shown in Fig.\,\ref{fig:d_and_js}.\\
\indent We focus on the case of the second-order phase transition. Substituting Eq.\,(\ref{eq:Psi_js}) into Eq.\,(\ref{eq:E_func}) with $c_6=c_6'=0$, the energy functional for Eq.\,(\ref{eq:Psi_js}) can be calculated as
\begin{eqnarray}
{\cal K} (\{ {\bf \Psi}_{\bm{p},\beta} \})
= -\frac{ V}{4c_4}(\varepsilon_p\sin^2\beta -|C_2|)^2, \label{eq:energy_functional}
 \end{eqnarray}
where $V= N_{\rm s} a^d$ is the volume of the system. Figure \ref{fig:energy} shows ${\cal K}(\{ {\bf \Psi}_{\bm{p},\beta} \})$ as a function of $\beta$ and the configuration of the $d$ vector (\ref{d_}) for each value of $\beta$.
 We find that the degeneracy of the energy functional with respect to $\beta$ due to $S^2_{\alpha,\beta}$ is lifted by a spin current: $\beta=0$ and $\pi$ correspond to the energy minima and $\beta=\pi/2$ to the maximum. In the former, since $R_z(\bm{p}\cdot \bm{r})$ does not change the $d$ vector that is in parallel with the $z$-axis, no spin currents are induced and thus ${\cal K}(\{ {\bf \Psi}_{\bm{p},\beta} \})$ takes the minima. In the latter, since the $d$ vector is in the $xy$ plane, spatial variation of the $d$ vector by $R_z(\bm{p}\cdot \bm{r})$ is maximized and thus  ${\cal K}(\{ {\bf \Psi}_{{\bm{p}},\beta}\})$ takes the maximum.
 The dynamical instability at $\beta=\pi/2$, therefore, occurs because it corresponds to an energetically unstable point.
 Even infinitesimally small fluctuations of the order parameter around $\beta=\pi/2$ grow and drive the system away from $\beta=\pi/2$ towards the stable states at $\beta=0$ or $\pi$. \\
\indent We note that the energy functional for the first-order phase transition also exhibits the maximum at $\beta=\pi/2$ and minima at $\beta=0$ and $\pi$. The origin of the dynamical instability is, therefore, the same as explained above. 
\par 
Our results do not exclude the possibility of stable spin currents in general. 
We expect that spin currents can be stabilized by the quadratic Zeeman effect.
In fact, stable spin currents have been observed in a spin-1 antiferromagnetic Bose-Einstein condensate (BEC) of $^{23}$Na atoms in the presence of the quadratic Zeeman effect \cite{kim17}, despite the fact that spin currents are predicted to be unstable in this system without the quadratic Zeeman effect \cite{fujimoto12,zhu15}.
\par
In the above argument, we have revealed that the instability of spin currents originates from the fact that the energy functional takes its maximum for a superfluid state carrying a spin current (see Fig.\,\ref{fig:energy}) that results in the dynamical instability of the spin nematic mode $\omega_\beta$.
The spin nematic mode $\omega_\beta$ involves small population of $m=0$ hyperfine state as indicated in Eq.~(\ref{eq:db}). In the presence of the quadratic Zeeman effect, since populating $m=0$ state from the polar state in Eq.~(\ref{eq:Psi_static}) costs finite energy due to the quadratic Zeeman effect, it is expected that the spin nematic mode may be stabilized and the energy functional takes minimum for a state carrying a spin current. 
However, it is beyond the scope of the paper to confirm this possibility using the TDGL equation, because it is necessary to derive the TDGL equation taking into account the quadratic Zeeman effect from scratch. Note that the insulating state in the presence of the quadratic Zeeman effect is different from the spin-singlet state in Eq.~(\ref{spin-singlet-mi}). 
We examine the possibility of stabilizing spin currents by the quadratic Zeeman effect using the discrete Gross-Pitaevskii (DGP) equation, which is valid in the weakly-interacting regime deep in the superfluid phase.
\par 
The spin-1 Bose-Hubbard model with the quadratic Zeeman effect \cite{deforgesdeparny18} is given by
\begin{eqnarray}
\hat{H}_q 
= 
\hat{H}-q \sum_{i,\alpha,\beta} \hat{b}_{i\alpha}^\dagger (F_z^2)_{\alpha\beta} \hat{b}_{i\beta},\label{Hq}
\end{eqnarray}%
where $q>0$. In the weakly interacting regime ($t\gg U_0$),
introducing the mean-field $\psi_{i\alpha} = \langle \hat{b}_{i\alpha}\rangle$, we obtain the energy functional 
\begin{eqnarray}
\langle \hat{H}_q \rangle 
&=& 
-q \sum_i (|\psi_{i1}|^2 + |\psi_{i-1}|^2)
- t \sum_{\langle i,j\rangle ,\alpha} (\psi_{i\alpha}^* \psi_{j\alpha} + {\rm c.c.})
\nonumber \\
&-& 
\mu \sum_{i,\alpha}|\psi_{i\alpha}|^2
+ \frac{U_0}{2} \sum_{i}(\sum_{\alpha}|\psi_{i\alpha}|^2)^2
\nonumber \\
&+& \frac{U_2}{2} \sum_{i}
(\sum_{\alpha,\beta}\psi_{i\alpha}^* \bm{F}_{\alpha\beta} \psi_{i\beta})^2.
\label{<Hq>}
\end{eqnarray}%
From $i\partial_t \psi_{i\alpha} = \delta \langle \hat{H}_q \rangle / \delta \psi_{i\alpha}^*$, the discrete Gross-Pitaevskii (DGP) equation can be derived as 
\begin{eqnarray}
i\partial_t \psi_{i\alpha} 
&=&
-t \sum_{j \in  V(i)} \psi_{j\alpha}
- q \psi_{i\alpha}(\delta_{\alpha,1}+ \delta_{\alpha,-1} )
\nonumber \\
&-&
\mu \psi_{i\alpha}
+ U_0 \psi_{i\alpha} \sum_{\beta}|\psi_{i\beta}|^2 
\nonumber \\
&+& 
U_2 \sum_{\beta,\gamma,\delta}
\psi_{i\beta}^* \bm{F}_{\beta\gamma} \psi_{i\gamma} \cdot \bm{F}_{\alpha\delta} \psi_{i\delta},\label{DGP}
\end{eqnarray}%
where $j \in V(i)$ denotes the summation over the nearest neighboring sites of site $i$. 
We consider the static solution that corresponds to Eq.~(80)
\begin{eqnarray}
\vec{\psi}_{i}^{0}(\beta) 
= 
\sqrt{n_{\rm c}}
\left[
\begin{array}{c}
-\frac{e^{-i \bm{p} \cdot \bm{r}_i}}{\sqrt{2}}\sin\beta \\
\cos \beta \\
\frac{e^{i\bm{p} \cdot \bm{r}_i}}{\sqrt{2}} \sin \beta
\end{array}
\right].\label{pb}
\end{eqnarray}%
Setting $\beta=\pi/2$, Eq.~(\ref{pb}) reduces to
\begin{eqnarray}
\vec{\psi}_{i}^{\rm 0}(\pi/2) 
= 
\sqrt{\frac{n_{\rm c}}{2}}
\left[
\begin{array}{c}
-e^{-i \bm{p} \cdot \bm{r}_i} \\
0 \\
e^{i\bm{p} \cdot \bm{r}_i}
\end{array}
\right], \label{pb_}
\end{eqnarray}%
which corresponds to Eq.~(72). We introduce fluctuation of the order parameter around $\vec{\psi}_{i}^{\rm 0}(\pi/2)$ as 
\begin{eqnarray}
\vec{\psi}_i = \vec{\psi}_{i}^{0}(\pi/2)
+ \delta \vec{\psi}_i,\label{aa}
\end{eqnarray}%
where 
\begin{eqnarray}
\delta \vec{\psi}_i 
= 
\left[
\begin{array}{c}
\delta \psi_{i1}e^{-i\bm{p} \cdot \bm{r}_i}\\
\delta \psi_{i0}\\
\delta \psi_{i-1}e^{i\bm{p} \cdot \bm{r}_i}
\end{array}
\right]. 
\end{eqnarray}%
Substituting Eq.~(\ref{aa}) into the DGP equation (\ref{DGP}) and linearizing with respect to fluctuation, we obtain
\begin{eqnarray}
&&\omega 
\left[
\begin{array}{c}
\varphi_{\bm k,0}\\
\varphi_{\bm k,0}^*
\end{array}
\right]
\nonumber \\
&=&
\left[
\begin{array}{cc}
\epsilon_{\bm k} -\epsilon_{\bm p} + U_2 n_{\rm c} -q & - U_2 n_{\rm c} \\
U_2 n_{\rm c} &  -(\epsilon_{\bm k}-\epsilon_{\bm p}  +  U_2 n_{\rm c} -q)
\end{array}
\right]
\left[
\begin{array}{c}
\varphi_{\bm k,0}\\
\varphi_{\bm k,0}^*
\end{array}
\right],\label{ph}\nonumber \\
\end{eqnarray}%
where $\varphi_{\bm{k},0}$ is the Fourier component of $\delta \psi_{i0}$ and $\epsilon_{\bm k}$ is given in Eq.~(C22). 
Solving Eq.~(\ref{ph}), we obtain the dispersion relation for the spin nematic mode corresponding to $\omega_\beta$ as
\begin{eqnarray}
\tilde{\omega}_\beta(\bm k,\bm p)
= \sqrt{(\epsilon_{\bm k}-\epsilon_{\bm p} + U_2 n_{\rm c} +q)^2 -(U_2n_{\rm c})^2}. \label{tt}
\end{eqnarray}%
Without the quadratic Zeeman effect ($q=0$), $\tilde{\omega}_\beta(\bm k,\bm p)$ becomes pure imaginary at $\bm k={\bf 0}$ for finite $\bm p$, which means that spin currents are dynamically unstable.
This demonstrates that the absence of spin supercurrents is not restricted within the TDGL equation, which is valid in the vicinity of the Mott insulating (MI) phase. Our results suggest that spin supercurrents are unstable in the entire superfluid phase without the quadratic Zeeman effect.
\par
In the presence of the quadratic Zeeman effect ($q>0$), one finds that $\tilde{\omega}_\beta$ does not exhibit dynamical instability for $\bm p$ smaller than the critical momentum $\bm p_{\rm c}$.
For small $\bm p$ ($pa\ll1$), the critical momentum $p_{\rm c}$ for the onset of the dynamical instability is obtained as
\begin{eqnarray}
 p_{\rm c} a = \sqrt{q/t}. \label{q/t}
\end{eqnarray}%
The quadratic Zeeman effect thus stabilizes spin currents. 
\par
Substituting Eq.~(\ref{pb}) into the energy functional (\ref{<Hq>}), we obtain 
\begin{eqnarray}
\frac{\langle \hat{H}_q \rangle}{n_{\rm c} N_{\rm s}} 
= 
\frac{\epsilon_{\bm p}}{2} -zt \cos^2 \beta -\mu - \frac{q}{2}\sin^2\beta.\label{Hqb}
\end{eqnarray}%
For small $\bm p$, Eq.~(\ref{Hqb}) reduces to
\begin{eqnarray}
\frac{\langle \hat{H}_q \rangle}{n_{\rm c} N_{\rm s}} 
=
-\mu - zt - \frac{ta^2}{2}(p_{\rm c}^2-p^2) \sin^2 \beta.\label{Hqbb} 
\end{eqnarray}%
Equation~(\ref{Hqbb}) shows that $\beta = \pi/2$ corresponds to an energy minimum for $p<p_{\rm c}$ and therefore the static solution (\ref{pb_}) carrying a finite spin current is stable as expected.
\section{conclusion}\label{sec:conclusion}

To summarize, we have studied the stability of supercurrents in the polar phase of antiferromagnetically interacting spin-1 bosons in an optical lattice using the TDGL equation.
We have calculated the critical momenta for supercurrents in the vicinity of the MI phase with even filling factors. 
We found that the critical momentum for mass currents is finite throughout the metastable SF phase, 
which demonstrates the superfluidity of the metastable SF state.
The critical momentum for spin currents was found to be zero.
We have found that this instability of spin currents originates from the fact that 
the polar state with a spin current corresponds to an energy saddle point, 
thereby an infinitesimal spin current causes a dynamical instability.\\
\indent Although we set $U_2/U_0$ optimal values for validity of the theory that is not realistic in the present experimental status, the obtained results may be applicable to $^{23}{\rm Na}$ with $U_2/U_0=0.04$. The first-order phase transition dominates the larger part of the phase boundary and the region of the metastable SF phase in the phase diagram gets larger for smaller $U_2/U_0$ \cite{kimura05,yamamoto13}. The effect of the first-order phase transition, therefore, may be remarkable and easy to study experimentally for $^{23}{\rm Na}$.\\
\indent Our predictions for the stability of supercurrents can be verified experimentally with current technologies. As for mass currents, we propose a similar setup with Ref.\,\cite{mun07}, where a mass current is induced by a moving optical lattice. Spin currents can be induced by applying a linear magnetic field gradient that induces counter flow of two spin components as in the experiment for spinor gases \cite{kim17}.

\acknowledgments
We thank R. Asaoka, I. Danshita, D. Kagamihara, Y. Kawaguchi, T. Kobayashi, and  D. Yamamoto for helpful discussions. The work of RY is supported by the Japanese Society for the Promotion of Science Grant-in-Aid for Scientific Research (KAKENHI Grants No.\,19K14616 and No.\,20H01838). This work is supported by the Japanese Society for the Promotion of Science Grant-in-Aid for Scientific Research (KAKENHI Grant No. 19K03691).
\appendix
\section{MATRIX ELEMENTS OF CREATION AND ANNIHILATION OPERATORS}
We present the details of the derivation of the matrix elements of creation and annihilation operators summarized in Table I that are necessary for the calculation of the ground-state energy (\ref{free}). The following derivation is based on Ref.\,\cite{mohamed13}.\\
\indent If one operates $\hat{b}_\alpha^\dagger\ (\alpha=0,\pm1)$ on the state $|S,m,n\rangle $, the magnetic quantum number $m$ and particle number $n$ change as $m\rightarrow m+\alpha$ and $n\rightarrow n+1$. Meanwhile, the total spin quantum number $S$ changes as $S\rightarrow S \pm 1$ by adding a spin-1 particle.
Thus, $\hat{b}_\alpha^\dagger|S,m,n\rangle$ can be written as
\begin{eqnarray}
\hat{b}_\alpha^\dagger|S,m,n\rangle
&=&M_{\alpha,S,m,n}|S+1,m+\alpha,n+1\rangle \nonumber \\
&&+N_{\alpha,S,m,n}|S-1,m+\alpha,n+1\rangle.
\end{eqnarray}
Analogously, $\hat{b}_\alpha | S, m, n \rangle $  can be written as
\begin{eqnarray}
\hat{b}_\alpha|S,m,n\rangle
&=&O_{\alpha,S,m,n}|S+1,m-\alpha,n-1\rangle\nonumber \\
&&+P_{\alpha,S,m,n}|S-1,m-\alpha,n-1\rangle.
\end{eqnarray}
We derive the recurrence formulas for $M_{\alpha,S,m,n},\ N_{\alpha,S,m,n},\ O_{\alpha,S,m,n},$ and $P_{\alpha,S,m,n}$.
From $-S\leq m\leq S$, they satisfy
\begin{eqnarray}
N_{1,S,S,n}=N_{0,S,S,n}=P_{0,S,S,n}=P_{-1,S,S,n}=0.
\end{eqnarray}
Then, we obtain
\begin{eqnarray}
\hat{b}_1^\dagger|S,S,n\rangle=M_{1,S,S,n}|S+1,S+1,n+1\rangle,\label{lll}
\end{eqnarray}
where $M_{1,S,S,n}$ is given as \cite{tsuchiya04}
\begin{eqnarray}
M_{1,S,S,n}=\sqrt{\frac{(S+1)(n+S+3)}{2S+3}}.
\end{eqnarray}
Other matrix elements can be derived by operating the ladder operators $\hat{S}^{\pm}=\hat{S}_x\pm i\hat{S}_y$. By simple calculations, one can show 
\begin{eqnarray*}
&&[\hat{S}^+,\hat{b}_1]=-\sqrt{2}\hat{b}_0,\ [\hat{S}^+,\hat{b}_0]=-\sqrt{2}\hat{b}_{-1},\ [\hat{S}^+,\hat{b}_{-1}]=0,\\
 &&[\hat{S}^-,\hat{b}_1]=0,\ [\hat{S}^-,\hat{b}_0]=-\sqrt{2}\hat{b}_{-1},\ [\hat{S}^-,\hat{b}_{-1}]=-\sqrt{2}\hat{b}_0,\\
 &&[\hat{S}_z,\hat{b}_1]=-\hat{b}_1,\ [\hat{S}_z,\hat{b}_0]=0,\ [\hat{S}_z,\hat{b}_{-1}]=\hat{b}_{-1},\\
 &&[\hat{b}_1,\hat{\Theta}^\dagger]=-2\hat{b}^\dagger_{-1},\ [\hat{b}_0,\hat{\Theta}^\dagger]=2\hat{b}_0^\dagger,\ [\hat{b}_{-1},\hat{\Theta}^\dagger]=-2\hat{b}_1^\dagger.
\end{eqnarray*}
Using the above commutation relations, we obtain
\begin{eqnarray}
&&\hat{S}^+\hat{b}_1^\dagger|S,m,n\rangle\nonumber  \\
&=&\sqrt{(S-m)(S+m+3)}M_{1,S,m,n}|S+1,m+2,n+1\rangle\nonumber\\
&+&\sqrt{(S-m-2)(S+m+1)}N_{1,S,m,n}|S-1,m+2,n+1\rangle\nonumber \\ 
\label{eq:sb} \\
&=&\hat{b}_1^\dagger\hat{S}^+|S,m,n\rangle\nonumber\\
&=&\sqrt{(S-m)(S+m+1)}M_{1,S,m+1,n}|S+1,m+2,n+1\rangle\nonumber\\
&+&\sqrt{(S-m)(S+m+1)}N_{1,S,m+1,n}|S-1,m+2,n+1\rangle.\nonumber \\ \label{eq:bs}
\end{eqnarray}
Comparing Eqs.\,(\ref{eq:sb}) and (\ref{eq:bs}), we obtain the recurrence formula for $M_{1, S,m,m}$ as 
\begin{eqnarray}
M_{1,S,m,n}=\sqrt{\frac{S+m+1}{S+m+3}}M_{1,S,m+1,n}.
\end{eqnarray}
In the same manner, we can derive other recurrence formulas
\begin{eqnarray}
M_{0,S,m,n}&=&\sqrt{\frac{(S-m)(S+m+1)}{(S-m+1)(S+m+2)}}M_{0,S,m+1,n}\nonumber \\
&&+\sqrt{\frac{2}{(S-m+1)(S+m+2)}}M_{1,S,m,n},\label{formulae_}\\
M_{-1,S,m,n}&=&\sqrt{\frac{(S-m)(S+m+1)}{(S-m+2)(S+m+1)}}M_{-1,S,m+1,n}\nonumber \\
&&+\sqrt{\frac{2}{(S-m+2)(S+m+1)}}M_{0,S,m,n},\nonumber \\ \\
N_{-1,S,m,n}&=&\sqrt{\frac{S(S+1)-m(m-1)}{S(S-1)-(m-1)(m-2)}}N_{-1,S,m-1,n},\nonumber \\ \\
N_{0,S,m,n}&=&\sqrt{\frac{S(S+1)-m(m-1)}{S(S-1)-m(m-1)}}N_{-1,S,m-1,n}\nonumber \\
&&+\sqrt{\frac{2}{S(S-1)-m(m-1)}}N_{-1,S,m,n},\\
N_{1,S,m,n}&=&\sqrt{\frac{S(S+1)-m(m-1)}{S(S-1)-m(m+1)}}N_{1,S,m-1,n}\nonumber \\
&&+\sqrt{\frac{2}{S(S+1)-m(m-1)}}N_{0,S,m,n}.\label{formulae}
\end{eqnarray}
Here, $M_{\alpha,S,S,n}$ and $N_{\alpha,S,S,n}$ satisfy
\begin{eqnarray}
n&=&\sum_\alpha\langle S,S,n|\hat{b}_\alpha^\dagger\hat{b}_{\alpha}|S,S,n\rangle\nonumber\\
&=&\sum_{\alpha}[M_{\alpha,S,S,n}^2+N_{\alpha,S,S,n}^2]-3.\label{eq:M_and_N} 
\end{eqnarray}
From Eq.\,(\ref{eq:M_and_N}) and $N_{1,S,S,n}=N_{0,S,S,n}=0$, the initial value of $N_{\alpha,S,m,n}$ is given by 
\begin{eqnarray}
N_{-1,S,S,n}=-\sqrt{3+n-\sum_{\alpha}M^2_{\alpha,S,S,n}}. \label{N-}
\end{eqnarray}
The sign of $N_{-1,S,S,n}$ is set to be consistent with Ref.\,\cite{tsuchiya04}. Solving the recurrence formulas (\ref{formulae_})-(\ref{formulae}) and (\ref{N-}), one obtains Table \ref{table} \cite{axel13}.\\
\indent One can calculate $O_{\alpha,S,m,n}$ and $P_{\alpha,S,m,n}$ as \cite{mohamed13}
\begin{eqnarray}
\langle S,m,n|\hat{b}^\dagger_\alpha|S-1,m-\alpha,n\rangle
&=&M_{\alpha,S-1,m-\alpha,n-1}\nonumber \\
&=&P_{\alpha,S,m,n},\\
\langle S,m,n|\hat{b}^\dagger_\alpha|S+1,m-\alpha,n\rangle
&=&M_{\alpha,S+1,m-\alpha,n-1}\nonumber \\
&=&O_{\alpha,S,m,n}.
\end{eqnarray}

\section{PERTURBATIVE MEAN-FIELD CALCULATION}
In this appendix, we present the details of the perturbative mean-field calculation in Sec.\,\ref{sec:perturbaticve_mean-field_theory}.
Since the ground-state energy should be invariant under the ${\rm U}(1)$ gauge transformation and spin rotations, it is convenient to expand the ground-state energy by the $d$ vector that transforms as a vector under spin rotations as 
\begin{eqnarray}
E&=&E^{(2)}+E^{(4)}+E^{(6)},\\
E^{(2)}&=&C_2(\bm{d}^\dagger\bm{d})=C_2n_{\rm c},\label{E2}\\
E^{(4)}&=&D_4(\bm{d}^\dagger\bm{d})^2+D_4'|\bm{d}\cdot\bm{d}|^2\nonumber \\
&=&C_4n_{\rm c}^2+C_4'n_{\rm c}^2\langle \bm{F}\rangle^2,\label{E4}\\
E^{(6)}&=&D_6(\bm{d}^\dagger\bm{d})^3+D_6'|\bm{d}\cdot\bm{d}|^2(\bm{d}^\dagger\bm{d})\nonumber \\
&=&C_6n_{\rm c}^3+C_6'n_{\rm c}^3\langle \bm{F}\rangle^2\label{E6},
\label{gakuhi}
\end{eqnarray}
where $C_m=D_m+D_m'$, $C_m'=-D_m'\ (m=4,6)$. Note that $\bm{d}^\dagger \bm{d}=n_{\rm c},\ \bm{d}\cdot \bm{d} = n_{\rm c}(\zeta_0^2 - 2\zeta_1 \zeta_{-1}),$ and $| \bm{d} \cdot \bm{d} |^2 = n_{\rm c}^2 (1-\langle \bm{F}\rangle^2)$.
To reduce the number of terms that need to be calculated, we set $\zeta_0=0$ in Eqs.\,(\ref{E2}), (\ref{E4}), (\ref{E6}), and (\ref{hopping}). We thus obtain
\begin{eqnarray}
E^{(2)}&=&C_2n_{\rm c}(|\zeta_1|^2+|\zeta_{-1}|^2),\\
E^{(4)}&=&D_4n_{\rm c}^2(|\zeta_1|^4+|\zeta_{-1}|^4)\nonumber \\
&&+(2D_4+4D_4')n_{\rm c}^2|\zeta_1|^2|\zeta_{-1}|^2, \\
E^{(6)}&=&D_6n_{\rm c}^3(|\zeta_1|^6+|\zeta_{-1}|^6)\nonumber\\
&&+(3D_6+4D_6')n_{\rm c}^3|\zeta_1|^2|\zeta_{-1}|^2(|\zeta_1|^2+|\zeta_{-1}|^2),\nonumber \\ \\ 
\hat{V}&=&-zt(\psi_1\hat{b}^\dagger_1+\psi_{-1}\hat{b}^\dagger_{-1}+\mathrm{h.c.}).
\end{eqnarray}
We employ the formulas of the standard perturbation theory \cite{messiah99}:
\begin{widetext}
\begin{eqnarray}
E^{(2)}&=&-\sum_{\alpha\neq g}\frac{|\langle \alpha|\hat{V}^\mathrm{pert}|g\rangle|^2}{\Delta E_{\alpha}},\label{eq:E2}\\
E^{(4)}&=&-\sum_{\alpha\beta\gamma \neq g}\langle g|\hat{V}^\mathrm{pert}|\alpha\rangle
\frac{\langle \alpha|\hat{V}^\mathrm{pert}|\beta\rangle}{\Delta E_{\alpha}}
\frac{\langle \beta|\hat{V}^\mathrm{pert}|\gamma\rangle}{\Delta E_{\beta}}
\frac{\langle \gamma|\hat{V}^\mathrm{pert}|g\rangle}{\Delta E_{\gamma}} -E^{(2)}\sum_{\alpha\neq g}\frac{|\langle \alpha|\hat{V}^\mathrm{pert}|g\rangle|^2}{\Delta E_{\alpha}^2},\label{eq:E4}\\
E^{(6)}&=&-\sum_{\alpha\beta\gamma\delta\epsilon\neq g}
\langle g|\hat{V}^\mathrm{pert}|\alpha\rangle
\frac{\langle \alpha|\hat{V}^\mathrm{pert}|\beta\rangle}{\Delta E_{\alpha}}
\frac{\langle \beta|\hat{V}^\mathrm{pert}|\gamma\rangle}{\Delta E_{\beta}}
\frac{\langle \gamma|\hat{V}^\mathrm{pert}|\delta\rangle}{\Delta E_{\gamma}}
\frac{\langle \delta|\hat{V}^\mathrm{pert}|\epsilon\rangle}{\Delta E_{\delta}}
\frac{\langle \epsilon|\hat{V}^\mathrm{pert}|g\rangle}{\Delta E_{\epsilon}} \nonumber \\
&+&\sum_{\alpha\beta\gamma\delta\neq g}
\langle g|\hat{V}^\mathrm{pert}|\alpha\rangle
\frac{\langle \alpha|\hat{V}^\mathrm{pert}|\beta\rangle}{\Delta E_{\alpha}}
\frac{\langle \beta|\hat{V}^\mathrm{pert}|\gamma\rangle}{\Delta E_{\beta}}
\frac{\langle \gamma|\hat{V}^\mathrm{pert}|\delta\rangle}{\Delta E_{\gamma}}
\frac{\langle \delta|\hat{V}^\mathrm{pert}|g\rangle}{\Delta E_{\delta}} \nonumber \\
&&\left(
\frac{1}{\Delta E_\alpha}
+\frac{1}{\Delta E_\beta}
+\frac{1}{\Delta E_\gamma}
+\frac{1}{\Delta E_\delta}
\right) 
-\sum_{\alpha\neq g}\frac{|\langle \alpha|\hat{V}^\mathrm{pert}|g\rangle|^6}{\Delta E_{\alpha}^5},\label{eq:E6}
\end{eqnarray}
where $|g \rangle$ is an unperturbative state, $\hat{V}^{\rm pert}$ represents the perturbation, $|\alpha\rangle,\,|\beta\rangle,\,|\gamma\rangle,\,|\delta\rangle,$ and $|\epsilon\rangle$ are the intermediate states, and $\Delta E_\alpha,\ \Delta E_\beta,\ \Delta E_\gamma,\ \Delta E_\delta$, and $\Delta E_\epsilon$ denote the excitation energy for $|\alpha\rangle,\ |\beta\rangle,\ |\gamma\rangle,\  |\delta\rangle,$ and $|\epsilon\rangle$ from $|g\rangle$.
Using the formulas (\ref{eq:E2}), (\ref{eq:E4}), (\ref{eq:E6}), and Table \ref{table}, the coefficients $C_2,\ D_4,\ D_4',\ D_6,$ and $D_6'$ can be calculated as 
\begin{eqnarray}
C_2&=&(zt)^2-\frac{(zt)^2}{3}
\left[
\frac{n+3}{\Delta E_{1,n+1}}+\frac{n}{\Delta E_{1,n-1}}
\right]\label{C-2},\\
 D_4&=&\frac{(zt)^4}{9}
  \left[  -\frac{6}{5}\frac{n(n-2)}{\Delta E_{1,n-1}^2 \Delta E_{2,n-2}}
   -\frac{1}{5}\frac{n(n+3)}{\Delta E_{2,n}}\left( \frac{1}{\Delta E_{1,n+1}}+\frac{1}{\Delta E_{1,n-1}}\right)^2 \right.\nonumber \\
  &-&\left. \frac{6}{5}\frac{(n+3)(n+5)}{\Delta E_{1,n+1}^2\Delta E_{2,n+2}}
  +\left( \frac{n+3}{\Delta E_{1,n+1}}+\frac{n}{\Delta E_{1,n-1}} \right)
  \left(\frac{n+3}{\Delta E_{1,n+1}^2}+\frac{n}{\Delta E_{1,n-1}^2}  \right)
   \right] \label{C_4__},
   \end{eqnarray}
   \begin{eqnarray}
  D_4'&=&-\frac{(zt)^4}{9}\left[  \frac{n(n+1)}{\Delta E_{1,n-1}^2\Delta E_{0,n-2}}
  + \frac{(n+2)(n+3)}{\Delta E_{1,n+1}^2\Delta E_{0,n+2}}
  -\frac{2}{5}\frac{n(n-2)}{\Delta E_{1,n-1}^2 \Delta E_{2,n-2}} \right. \nonumber\\
  &+& \frac{3}{5}\frac{n(n+3)}{\Delta E_{2,n}}
  \left( \frac{1}{\Delta E_{1,n+1}}+\frac{1}{\Delta E_{1,n-1}} \right)^2
  -\left. \frac{2}{5}\frac{(n+3)(n+5)}{\Delta E_{1,n+1}^2\Delta E_{2,n+2}}\right]\label{C_4'__},
  \end{eqnarray}
  \begin{eqnarray*}
 D_6
  &=&
  \left[
  -\frac{2}{35}\frac{(n+7)(n+5)(n+3)}{\Delta E_{3,n+3}\Delta E_{2,n+2}^2\Delta E_{1,n+1}^2}
  \right.
  -\frac{2}{35}\frac{(n-4)(n-2)n}{\Delta E_{3,n-3} \Delta E_{2,n-2}^2 \Delta E_{1,n-1}^2 }\\
  &-&\frac{2}{525}\frac{n(n+3)(n+5)}{\Delta E_{3,n+1}}
  \left(
  \frac{1}{\Delta E_{2,n+2} \Delta E_{1,n+1}}
  +\frac{1}{\Delta E_{2,n} \Delta E_{1,n+1}}
  +\frac{1}{\Delta E_{2,n}\Delta E_{1,n-1}}
  \right)^2 \\
  &-&\frac{2}{525}\frac{n(n-2)(n+3)}{\Delta E_{3,n-1}}
  \left(
  \frac{1}{\Delta E_{2,n-2} \Delta E_{1,n-1}}
  +\frac{1}{\Delta E_{2,n} \Delta E_{1,n+1} }
  +\frac{1}{\Delta E_{2,n}\Delta E_{1,n-1}}
  \right)^2 \\
   &-&\frac{1}{3}\frac{(n+3)}{\Delta E_{1,n+1}}
  \left(
  \frac{2(n+5)}{5}\frac{1}{\Delta E_{2,n+2}\Delta E_{1,n+1}}
  +\frac{n}{15}\frac{1}{\Delta E_{2,n}\Delta E_{1,n+1}}
  +\frac{n}{15}\frac{1}{\Delta E_{2,n}\Delta E_{1,n-1}}
  \right)^2 \\
  &-&\frac{1}{3}\frac{n}{\Delta E_{1,n-1}}
  \left(
  \frac{(n+3)}{15}\frac{1}{\Delta E_{2,n}\Delta E_{1,n+1}}
  +\frac{(n+3)}{15}\frac{1}{\Delta E_{2,n}\Delta E_{1,n-1}}
  +\frac{2(n-2)}{5}\frac{1}{\Delta E_{2,n-2}\Delta E_{1,n-1}}
  \right)^2 \\
  &+&\left\{
  \frac{2}{15}\frac{(n+3)(n+5)}{\Delta E_{2,n+2}\Delta E_{1,n+1}^2}
  \right.
  +\frac{1}{45}\frac{n(n+3)}{\Delta E_{2,n}}
  \left(
  \frac{1}{\Delta E_{1,n+1}}
  +\frac{1}{\Delta E_{1,n-1}}
  \right)^2 \\
   &+&\left.
  \frac{2}{15}\frac{(n-2)n}{\Delta E_{2,n-2}\Delta E_{1,n-1}^2}
  \right\}
  \left(
  \frac{(n+3)}{3}\frac{1}{\Delta E_{1,n+1}^2}
  +\frac{n}{3}\frac{1}{\Delta E_{1,n-1}^2}
  \right) \\
  &+&\left\{
  \frac{4}{15}\frac{(n+3)(n+5)}{\Delta E_{2,n+2}\Delta E_{1,n+1}^3}
  \right.
  +\frac{2}{45}\frac{n(n+3)}{\Delta E_{2,n}}
  \left(
  \frac{1}{\Delta E_{1,n+1}^2}
  +\frac{1}{\Delta E_{1,n-1}^2}\right)\left(\frac{1}{\Delta E_{1,n+1}}
  +\frac{1}{\Delta E_{1,n-1}}
  \right) \\
  &+&\left.
  \frac{4}{15}\frac{(n-2)n}{\Delta E_{2,n-2}\Delta E_{1,n-1}^3}
  \right\}
  \left(
  \frac{(n+3)}{3\Delta E_{1,n+1}}+\frac{n}{3\Delta E_{1,n-1}}
  \right) \\
  &+&\left\{
  \frac{2}{15}\frac{(n+3)(n+5)}{\Delta E_{2,n+2}^2\Delta E_{1,n+1}^2}
  +\frac{1}{45}\frac{n(n+3)}{\Delta E_{2,n}^2}
  \left(
  \frac{1}{\Delta E_{1,n+1}}+\frac{1}{\Delta E_{1,n-1}}
  \right)^2
  \right. \\
  &+&\left.
  \frac{2}{15}\frac{(n-2)n}{\Delta E_{1,n-1}^2\Delta E_{2,n-2}^2}
  \right\}
  \left(
  \frac{(n+3)}{3\Delta E_{1,n+1}}+\frac{n}{3\Delta E_{1,n-1}}
  \right) \\
  &-&\left(
  \frac{(n+3)}{3\Delta E_{1,n+1}^3}+\frac{n}{3\Delta E_{1,n-1}^3}
  \right)
  \left(
  \frac{(n+3)}{3\Delta E_{1,n+1}}
  +\frac{n}{3\Delta E_{1,n-1}}
  \right)^2  \\
  &-&\left.
  \left(
  \frac{(n+3)}{3\Delta E_{1,n+1}}+\frac{n}{3\Delta E_{1,n-1}}
  \right)
  \left(
  \frac{n+3}{3\Delta E_{1,n+1}^2}
  +\frac{n}{3\Delta E_{1,n-1}^2}
  \right)^2
  \right](zt)^6, \\ 
  \end{eqnarray*}
  \begin{eqnarray*}
  D_6'&=&\frac{(zt)^6}{4}\left[
  -\frac{46}{525}\frac{n(n+3)(n+5)}{\Delta E_{3,n+1}}
  \left(
  \frac{1}{\Delta E_{1,n-1}\Delta E_{2,n}}
  +\frac{1}{\Delta E_{1,n+1}\Delta E_{2,n}}
  +\frac{1}{\Delta E_{1,n+1}\Delta E_{2,n+2}}
  \right)^2 \right.\\
    &-&\frac{46}{525}\frac{(n-2)n(n+3)}{\Delta E_{3,n-1}}
  \left(
  \frac{1}{\Delta E_{1,n-1}\Delta E_{2,n}}
  +\frac{1}{\Delta E_{1,n+1}\Delta E_{2,n}}
  +\frac{1}{\Delta E_{1,n-1}\Delta E_{2,n-2}}
  \right)^2 \\
    &-&\frac{18}{525}\frac{(n+3)(n+5)(n+7)}{\Delta E_{3,n+3}\Delta E_{1,n+1}^2\Delta E_{2,n+2}^2}
  -\frac{18}{525}\frac{(n-4)(n-2)n}{\Delta E_{3,n-3}\Delta E_{1,n-1}^2\Delta E_{2,n-2}^2} \\
   &-&\frac{1}{3}\frac{(n+2)(n+3)(n+5)}{\Delta E_{1,n+3}}
  \left(
  \frac{8}{15\Delta E_{2,n+2}\Delta E_{1,n+1}}
  +\frac{2}{3\Delta E_{0,n+2}\Delta E_{1,n+1}}
  \right)^2 \\
  &-&\frac{2}{3}\frac{(n+3)}{\Delta E_{1,n+1}}
  \left(
  \frac{7n}{15\Delta E_{2,n}\Delta E_{1,n+1}}
  +\frac{7n}{15\Delta E_{2,n}\Delta E_{1,n-1}}
  +\frac{2(n+5)}{15\Delta E_{2,n+2}\Delta E_{1,n+1}}
  +\frac{2(n+2)}{3\Delta E_{0,n+2}\Delta E_{1,n+1}}
  \right) \\
  &&
  \left(
  \frac{2(n+5)}{5\Delta E_{2,n+2}\Delta E_{1,n+1}}
  +\frac{n}{15\Delta E_{2,n}\Delta E_{1,n+1}}
  +\frac{n}{15\Delta E_{2,n}\Delta E_{1,n-1}}
  \right) \\
   &-&\frac{1}{3}\frac{n}{\Delta E_{1,n-1}}
  \left(
  \frac{7(n+3)}{15\Delta E_{2,n}\Delta E_{1,n+1}}
  +\frac{7(n+3)}{15\Delta E_{2,n}\Delta E_{1,n-1}}
  +\frac{2(n-2)}{15\Delta E_{2,n-2}\Delta E_{1,n-1}}
  +\frac{2(n+1)}{3\Delta E_{0,n-2}\Delta E_{1,n-1}}
  \right)^2 
   \end{eqnarray*}
  \begin{eqnarray*}
  &-&\frac{1}{3}\frac{(n+3)}{\Delta E_{1,n+1}}
  \left(
  \frac{2(n+5)}{15\Delta E_{2,n+2}\Delta E_{1,n+1}}
  +\frac{2(n+2)}{3\Delta E_{0,n+2}\Delta E_{1,n+1}}
  +\frac{7n}{15\Delta E_{2,n}\Delta E_{1,n-1}}
  +\frac{7n}{15\Delta E_{2,n}\Delta E_{1,n+1}}
  \right)^2 \\
  &-&\frac{2}{3}\frac{n}{\Delta E_{1,n-1}}
  \left(
  \frac{(n+3)}{15\Delta E_{2,n}\Delta E_{1,n+1}}
  +\frac{(n+3)}{15\Delta E_{2,n}\Delta E_{1,n-1}}
  +\frac{2(n-2)}{5\Delta E_{2,n-2}\Delta E_{1,n-1}}
  \right) \\
  &&\left(
  \frac{7(n+3)}{15\Delta E_{2,n}\Delta E_{1,n+1}}
  +\frac{7(n+3)}{15\Delta E_{2,n}\Delta E_{1,n-1}}
  +\frac{2(n-2)}{15\Delta E_{2,n-2}\Delta E_{1,n-1}}
  +\frac{2(n+1)}{3\Delta E_{0,n-2}\Delta E_{1,n-1}}
  \right) \\
    &-&\frac{1}{3}\frac{(n-2)n(n+1)}{\Delta E_{1,n-3}}
  \left(
  \frac{8}{15\Delta E_{2,n-2}\Delta E_{1,n-1}}
  +\frac{2}{3\Delta E_{0,n-2}\Delta E_{1,n-1}}
  \right)^2 \\
  &+&\left\{
  \frac{4}{9}\frac{n(n+1)}{\Delta E_{1,n-1}^2\Delta E_{0,n-2}}
  +\frac{4}{9}\frac{(n+2)(n+3)}{\Delta E_{1,n+1}^2\Delta E_{0,n+2}}
  +\frac{2}{9}\frac{n(n-2)}{\Delta E_{1,n-1}^2\Delta E_{2,n-2}}
  \right. \\
  &+&\frac{1}{3}\frac{n(n+3)}{\Delta E_{2,n}}
  \left(
  \frac{1}{\Delta E_{1,n+1}}
  +\frac{1}{\Delta E_{1,n-1}}
  \right)^2
  \left.
  +\frac{2}{9}\frac{(n+3)(n+5)}{\Delta E_{1,n+1}^2\Delta E_{2,n+2}}
  \right\}
  \left(
  \frac{(n+3)}{3\Delta E_{1,n+1}^2}
  +\frac{n}{3\Delta E_{1,n-1}^2}
  \right) \\
  &+&2\left\{
  \frac{4}{9}\frac{n(n+1)}{\Delta E_{1,n-1}^3\Delta E_{0,n-2}}
  +\frac{4}{9}\frac{(n+2)(n+3)}{\Delta E_{1,n+1}^3\Delta E_{0,n+2}}
  +\frac{2}{9}\frac{n(n-2)}{\Delta E_{1,n-1}^3\Delta E_{2,n-2}}
  +\frac{2}{9}\frac{(n+3)(n+5)}{\Delta E_{1,n+1}^3\Delta E_{2,n+2}}\right.\\
  &+&\frac{1}{3}\frac{n(n+3)}{\Delta E_{2,n}}
  \left(
  \frac{1}{\Delta E_{1,n+1}}
  +\frac{1}{\Delta E_{1,n-1}}
  \right)\left.
  \left(
  \frac{1}{\Delta E_{1,n+1}^2}
  +\frac{1}{\Delta E_{1,n-1}^2}
  \right)
  \right\}
  \left(
  \frac{(n+3)}{3\Delta E_{1,n+1}}
  +\frac{n}{3\Delta E_{1,n-1}}
  \right) \\
  &+&\left\{
  \frac{4}{9}\frac{n(n+1)}{\Delta E_{1,n-1}^2\Delta E_{0,n-2}^2}
  +\frac{4}{9}\frac{(n+2)(n+3)}{\Delta E_{1,n+1}^2\Delta E_{0,n+2}^2}
  +\frac{2}{9}\frac{n(n-2)}{\Delta E_{1,n-1}^2\Delta E_{2,n-2}^2} \right.\\
  &+&\frac{1}{3}\frac{n(n+3)}{\Delta E_{2,n}^2}
  \left.\left(
  \frac{1}{\Delta E_{1,n+1}}
  +\frac{1}{\Delta E_{1,n-1}}
  \right)^2
  +\frac{2}{9}\frac{(n+3)(n+5)}{\Delta E_{1,n+1}^2\Delta E_{2,n+2}^2}
  \right\}
  \left(
  \frac{(n+3)}{3\Delta E_{1,n+1}}
  +\frac{n}{3\Delta E_{1,n-1}}
  \right) \\
  &-&3\left(
  \frac{(n+3)}{3\Delta E_{1,n+1}^3}+\frac{n}{3\Delta E_{1,n-1}^3}
  \right)
  \left(
  \frac{(n+3)}{3\Delta E_{1,n+1}}
  +\frac{n}{3\Delta E_{1,n-1}}
  \right)^2  \\
  &-&3\left.\left(
  \frac{(n+3)}{3\Delta E_{1,n+1}}+\frac{n}{3\Delta E_{1,n-1}}
  \right)
  \left(
  \frac{n+3}{3\Delta E_{1,n+1}^2}
  +\frac{n}{3\Delta E_{1,n-1}^2}
  \right)^2\right]-\frac{3}{4}D_6.
\end{eqnarray*}
Here, $\Delta E_{S, n+a} = E^0(S,n+a) -  E^0(0,n)$. 
\end{widetext}
\section {DERIVATION OF THE TDGL EQUATION}  \label{app:tdgl}
In this Appendix, we summarize the derivation of the TDGL equation (\ref{tdgl}).
Using the coherent-state path integral \cite{fisher89}, the grand partition function for the spin-1 Bose-Hubbard model can be written as 
\begin{eqnarray} 
\Xi=\int\prod_{\alpha} \mathcal{D}b^*_{i\alpha} \mathcal{D}b_{i\alpha}e^{-S(\{b_{i\alpha}\})},
\end{eqnarray}
where the action $S(\{b_{i\alpha}\})$ is given by
\begin{eqnarray}
S(\{b_{i\alpha}\})
&=&\int^{\beta}_{0}{\rm d}\tau \left[ \sum_{i,\alpha}(b_{i\alpha}^*\partial_\tau b_{i\alpha}-\mu b_{i\alpha}^*b_{i\alpha})
\right.\nonumber\\
&&-\left.\sum_{ i,j,\alpha}t_{ij}b^*_{i\alpha}b_{j\alpha}+\frac{U_0}{2}\sum_{i,\alpha,\beta}b^*_{i\alpha}b^*_{i\beta}b_{i\beta}b_{i\alpha}\right.\nonumber\\
&&+\left. \frac{U_2}{2}\sum_{i,\alpha,\beta,\gamma,\delta} b_{i\alpha}^*b_{i\gamma}^* \bm{F}_{\alpha\beta} \cdot \bm{F}_{\gamma\delta}b_{i\beta}b_{i\delta} \right].\label{Ssbh}
\end{eqnarray} 
Here, the creation and annihilation operators in Eq.\,(\ref{BHM}) are replaced by the $c$-number field $b_{i\alpha}(\tau)$ that depends on imaginary time  $\tau$ and site $i$.
$\beta=1/T$ is the inverse temperature and $t_{ij}$ is an element of the hopping matrix $\hat{t}$.
For the nearest-neighbor hopping term, $t_{ij}$ is given as
\begin{eqnarray}
t_{ij}=\left\{
\begin{array}{l}
t,\ \ \mathrm{if\ {\it i}\ and\ {\it j}\ are\ nearest\ neighboring},\\
0,\ \ \mathrm{otherwise}.
\end{array}
\right.
\end{eqnarray}\\
\indent We introduce the auxiliary fields $\phi_{i\alpha} $ and  $\phi_{i\alpha}^*$ via the Hubbard-Stratonovich transformation \cite{negele98}:
\begin{eqnarray}
&&\int\mathcal{D}\phi^*_{i\alpha}\mathcal{D}\phi_{i\alpha}\exp\left[-\int {\rm d} \tau (\vec{\phi}_\alpha^\dagger-\vec{b}_\alpha^\dagger \hat{t})(\hat{t})^{-1}(\vec{\phi}_\alpha-\hat{t}\vec{b}_\alpha)\right]\nonumber \\
&&=\mathrm{const},\label{eq:H-S_t}
\end{eqnarray}
where $\vec{\phi}_\alpha=(\{\phi_{i\alpha}\})^T$ and $\vec{b}_\alpha=(\{b_{i\alpha}\})^T$. Multiplying Eq.\,(\ref{eq:H-S_t}) to Eq.\,(\ref{Ssbh}), we obtain
\begin{eqnarray}
\Xi =\int \prod_{\alpha}\mathcal{D}\phi_{i\alpha}^*\mathcal{D}\phi_{i\alpha}\prod_{\alpha}\mathcal{D}b^*_{i\alpha}\mathcal{D}b_{i\alpha}e^{-S(\{b_{i\alpha}\},\{\phi_{i\alpha}\})},\nonumber \\
\end{eqnarray}
where
\begin{eqnarray}
S
&=&
\int^{\beta}_{0}{\rm d}\tau
\sum_{i,j,\alpha} \phi_{i\alpha}^*(\hat{t}^{-1})_{ij}\phi_{j\alpha}
+S^0+S^\mathrm{pert},\label{eq:S}\\
S^0
&=&
\int^{\beta}_{0} {\rm d} \tau
\left[
	\sum_{i,\alpha}(b^*_{i\alpha}\partial_\tau b_{i\alpha}-\mu b^*_{i\alpha}b_{i\alpha})
\right.\nonumber\\
&&+\frac{U_0}{2}\sum_{i,\alpha,\beta}b^*_{i\alpha}b^*_{i\beta}b_{i\beta}b_{i\alpha} \nonumber \\
&&\left.
	+\frac{U_2}{2}\sum_{i,\alpha,\beta,\gamma,\delta}b^*_{i\alpha}b^*_{i\gamma} 
	\bm{F}_{\alpha\beta} \cdot\bm{F}_{\gamma\delta}b_{i\beta}b_{i\delta}
\right],\\
S^\mathrm{pert}
&=&
-\int^{\beta}_{0} {\rm d} \tau\sum_{i,\alpha}(\phi_{i\alpha}b^*_{i\alpha}+\mathrm{c.c.}).
\end{eqnarray}
Integrating out $b_{i\alpha}$ and $b^*_{i\alpha}$, we obtain
\begin{eqnarray}
\Xi = \int \prod_{\alpha} \mathcal{D}\phi_{i\alpha}^* \mathcal{D}\phi_{i\alpha}e^{-S^{\mathrm{eff}}(\{ \phi_{i\alpha}\})},
\end{eqnarray}
where the effective action $S^{\mathrm{eff}}$ is given by
\begin{eqnarray}
S^\mathrm{eff}
&=&
E^0(0,n)N_{\rm s}\beta -\log\langle e^{-S^\mathrm{pert}}\rangle_0 \nonumber \\
&&+\int^{\beta}_{0} {\rm d} \tau
\sum_{i,j,\alpha}(\hat{t}^{-1})_{ij}\phi^*_{i\alpha}\phi_{j\alpha}.\label{eq:Seff}
\end{eqnarray}
Here, $\langle \hat{\cal O} \rangle_0$ denotes the expectation value of the operator $\hat{\cal O}$ with respect to $S^0$: 
\begin{eqnarray}
\langle \hat{\mathcal O}\rangle_0
\equiv 
\frac{\int \prod_{\alpha}\mathcal{D}b_{i\alpha}\mathcal{D}b^*_{i\alpha}\hat{\mathcal{O}}e^{-S^0}}{\int \prod_{\alpha}\mathcal{D}b_{i\alpha}\mathcal{D}b^*_{i\alpha}e^{-S^0}}.
\end{eqnarray}
We perform the cumulant expansion in the second term in Eq.\,(\ref{eq:Seff}) as \cite{kubo62}
\begin{eqnarray}
\log \langle e^{-S^\mathrm{pert}}\rangle_0&\sim&\frac{1}{2!}\langle (S^\mathrm{pert})^2\rangle_{\rm c}+\frac{1}{4!}\langle (S^\mathrm{pert})^4\rangle_{\rm c}\nonumber \\
&&+\frac{1}{6!}\langle (S^\mathrm{pert})^6\rangle_{\rm c},\label{eq:cum}
\end{eqnarray}
where $\langle \hat{\mathcal O}\rangle_{\rm c}$ denotes the cumulant average of $\hat{\mathcal{O}}$ defined as
\begin{eqnarray}
\log \langle e^{\hat{\mathcal O}}\rangle_0=\sum_{n=1}^\infty \frac{\langle (\hat{\mathcal O})^n\rangle_{\rm c}}{n!}.
\end{eqnarray}
We note that, since Eq.\,(\ref{eq:S}) is invariant under the {\rm U}(1) gauge transformation
$b_{i\alpha}\rightarrow e^{i\theta}b_{i\alpha}$ and $\phi_{i\alpha}\rightarrow e^{i\theta}\phi_{i\alpha}$,
all the odd-order terms in the cumulant expansion (\ref{eq:cum}) vanish.
Each term in the right-hand side of Eq.\,(\ref{eq:cum}) can be written as  
\begin{eqnarray}
\langle(S^\mathrm{pert})^2\rangle_{\rm c}&=&\langle(S^\mathrm{pert})^2\rangle_0,\label{eq:cum2}\\
\langle(S^\mathrm{pert})^4\rangle_{\rm c}&=&\langle(S^\mathrm{pert})^4\rangle_0-3\langle(S^\mathrm{pert})^2\rangle_0^2,\label{eq:cum4}\\
\langle(S^\mathrm{pert})^6\rangle_{\rm c}&=&\langle(S^\mathrm{pert})^6\rangle_0-15\langle(S^\mathrm{pert})^2\rangle_0\langle(S^\mathrm{pert})^4\rangle_0\nonumber \\
&&+30\langle (S^\mathrm{pert})^2\rangle^3_0.\label{eq:cum6}
\end{eqnarray}\\
\indent To describe the critical behavior of the SF phase, it is sufficient to retain the first and second-order time-derivative terms \cite{fisher89}. Using Eqs.\,(\ref{eq:cum2})-(\ref{eq:cum6}) and taking the zero-temperature limit $\beta\rightarrow \infty$, we obtain
\begin{widetext}
\begin{eqnarray}
S^\mathrm{eff}
&\sim&
 \int^\infty_0 {\rm d} \tau
 \left\{
 	E^0(0,n)N_{\rm s} +
 	 \sum_{i,j,\alpha}(\hat{t}^{-1})_{ij}\phi_{i\alpha}^*\phi_{j\alpha}
\right.\nonumber \\
&+&\sum_{i,\alpha}\left[K\frac{\phi_{i\alpha}^*}{zt}\frac{\partial_\tau\phi_{i\alpha}}{zt}+J\left|\frac{\partial_\tau\phi_{i\alpha}^*}{zt}\right|^2
+(C_2-zt)\left|\frac{\phi_{i\alpha}}{zt}\right|^2+C_4\left|\frac{\phi_{i\alpha}}{zt}\right|^4+C_6\left|\frac{\phi_{i\alpha}}{zt}\right|^6\right]\nonumber\\
&+&\left.\sum_i\left[C_4'\left(\sum_{\alpha,\beta}\frac{\phi_{i\alpha}^*}{zt}\bm{F}_{\alpha\beta}\frac{\phi_{i\beta}}{zt}\right)^2
+C_6'\left(\sum_{\alpha,\beta}\frac{\phi_{i\alpha}^*}{zt}\bm{F}_{\alpha\beta}\frac{\phi_{i\beta}}{zt}\right)^2\sum_{\alpha}\left|\frac{\phi_{i\alpha}}{zt}\right|^2\right]\right\},\label{Seff}
\end{eqnarray}
\end{widetext}
where $K$ and $J$ are given by
\begin{eqnarray}
K=\frac{(zt)^2}{3}\left[\frac{n+3}{\Delta E_{1,n+1}^2}-\frac{n}{\Delta E_{1,n-1}^2}\right],\label{K-}\\
J=\frac{(zt)^2}{3}\left[\frac{n+3}{\Delta E_{1,n+1}^3}+\frac{n}{\Delta E_{1,n-1}^3}\right].
\end{eqnarray}
By the Fourier transform, the second term in Eq.\,(\ref{Seff}) can be written as
\begin{eqnarray}
\int^\infty_0 {\rm d} \tau  \sum_{i,j,\alpha}(\hat{t}^{-1})_{ij}\phi_{i\alpha}^*\phi_{j\alpha}
&=&
-\sum_{\bm{k},\omega,\alpha} \frac{|\bar{\phi}_{\alpha}(\bm{k},\omega)|^2}{\epsilon_{\bm k}}\nonumber\\ 
&\sim&  
\sum_{\bm{k},\omega,\alpha}\left| \frac{\bar{\phi}_\alpha(\bm{k},\omega)}{zt} \right|^2[ zt + (ka)^2 ],\nonumber \\\label{eq:kinetic}
\end{eqnarray}
where 
\begin{eqnarray}
\phi_{i\alpha}(\tau) &=& \sum_{\bm{k},\omega} \bar{\phi}_\alpha(\bm{k},\omega)e^{i ( \bm{k} \cdot \bm{r}_i -\omega \tau)},\\
\epsilon_{\bm{k}} &=& -2t\sum_{l=1}^d\cos(k_l a).\label{c22}
\end{eqnarray}
Here, $\bm{r}_i\equiv(x_ia,y_ia,z_ia)^T$ and $a$ is the lattice constant.
We have taken the long-wavelength limit, i.e., $ka\ll1$ in Eq.\,(\ref{eq:kinetic}).
In the continuum limit that is effective in the vicinity of the phase boundary, making replacements $\phi_{i\alpha}(\tau)/(a^{d/2}zt)\rightarrow \Psi_{\alpha}(\tau,\bm{r})$, $S^{\rm eff}$ reduces to the Ginzburg-Landau (GL) action $S^\mathrm{GL}$:
\begin{widetext}
\begin{eqnarray}
S^{\mathrm{GL}}&=&  \int^\infty_0 {\rm d} \tau \int {\rm d}^d x  
\left[ E^0(0,n)
+
K{\bf \Psi}^\dagger \partial_\tau {\bf \Psi}
+
J(\partial_\tau {\bf \Psi}^\dagger)(\partial_\tau {\bf \Psi})
+
\frac{1}{2m^*}(\nabla{\bf \Psi}^\dagger)\cdot(\nabla{\bf \Psi})
\right.\nonumber\\
&&\left.
+
C_2({\bf \Psi}^\dagger {\bf \Psi})
+
c_4({\bf \Psi}^\dagger{\bf \Psi})^2
+c_6({\bf \Psi}^\dagger{\bf \Psi})^3
+c_4'\langle \! \langle \bm{F} \rangle\! \rangle^2
+c_6'\langle \! \langle \bm{F}\rangle\! \rangle^2({\bf \Psi}^\dagger{\bf \Psi})
\right].\label{act:gl}
\end{eqnarray}
Here, 
$c_4 = a^d C_4,\ c_4' = a^d C_4', c_6 = a^{2d} C_6,\ c_6' = a^{2d}C_6',\ {\bf \Psi} = (\Psi_1,\Psi_0,\Psi_{-1})^T\ {\rm is\  the\ SF\ order\ parameter},\ \langle \! \langle \bm{F} \rangle \! \rangle = \sum_{\alpha,\beta} \Psi_{\alpha}^* \bm{F}_{\alpha\beta} \Psi_{\beta}\ {\rm is\ the\ spin\ average},$
and $m^*=1/(2ta^2)$ is the effective mass.
Setting $\partial_\tau \Psi_\alpha=0$ in Eq.\,(\ref{act:gl}), the energy functional for a static solution ${\bf \Psi}$ can be written as
\begin{eqnarray}
{\cal K} (\{ {\bf \Psi} \})=\int {\rm d}^d x 
&&\left[
	\frac{1}{2m^*} (\nabla {\bf \Psi}^\dagger) \cdot (\nabla {\bf \Psi})
	+ C_2({\bf \Psi}^\dagger {\bf \Psi})
	+ c_4 ({\bf \Psi}^\dagger  {\bf \Psi})^2
	+ c_6 ({\bf \Psi}^\dagger  {\bf \Psi})^3 
	+ c_4' \langle \! \langle {\bm F} \rangle \! \rangle^2
	+ c_6' \langle \! \langle {\bm F} \rangle\! \rangle^2({\bf \Psi}^\dagger  {\bf \Psi})
\right].\nonumber \\ \label{eq:E_func}
\end{eqnarray}
From $\frac{\delta S^{\rm GL}}{\delta \Psi_{\alpha}^*}=0$, we finally obtain the TDGL equation
\begin{eqnarray}
iK\partial_t\Psi_\alpha-J\partial_t^2\Psi_\alpha&=&
-\frac{\nabla^2}{2m^*}\Psi_\alpha
+
C_2\Psi_\alpha
+
2c_4({\bf \Psi}^\dagger {\bf \Psi})\Psi_\alpha
+
3c_6({\bf \Psi}^\dagger {\bf \Psi})^2\Psi_\alpha 
+
c_6'\langle \! \langle \bm{F} \rangle \! \rangle^2\Psi_\alpha\nonumber \\
&&
+
2c_4'\langle \! \langle \bm{F} \rangle \! \rangle\cdot\sum_\beta(\bm{F}_{\alpha\beta}\Psi_\beta) 
+
2c_6'({\bf \Psi}^\dagger {\bf \Psi})\langle \! \langle \bm{F} \rangle\!  \rangle\cdot\sum_\beta(\bm{F}_{\alpha\beta}\Psi_\beta)
.\label{tdgl}
\end{eqnarray}
Note that Eq.~(\ref{tdgl}) is valid for low energy dynamics, in which slow temporal variation of ${\bf \Psi}$ is allowed. If $K\neq0$, the first-order time-derivative term is dominant and the second-order one should be neglected.
\end{widetext}
\begin{table*}[]
\caption{Matrix elements}
\begin{tabular}{|c|c|cccccc|}
\cline{1-8}
$S$ & $m$  & $M_{1,S,m,n}$                     & $M_{0,S,m,n}$                & $M_{-1,S,m,n}$                    & $N_{1,S,m,n}$ & $N_{0,S,m,n}$ & $N_{-1,S,m,n}$ \\ \cline{1-8}
0 & 0  & $\sqrt{\frac{n+3}{3}}$            & $\sqrt{\frac{n+3}{3}}$       & $\sqrt{\frac{n+3}{3}}$            & 0             & 0             & 0              \\
1 & 1  & $\sqrt{\frac{2(n+4)}{5}}$         & $\sqrt{\frac{n+4}{5}}$       & $\sqrt{\frac{n+4}{15}}$           & 0              &  0             &$-\sqrt{\frac{n+1}{3}}$                \\
1 & 0  & $\sqrt{\frac{n+4}{5}}$            & $2\sqrt{\frac{n+4}{15}}$     & $\sqrt{\frac{n+4}{5}}$            &  0             & $\sqrt{\frac{n+1}{3}} $             &  0              \\
1 & -1 & $\sqrt{\frac{n+4}{15}}$           & $\sqrt{\frac{n+4}{5}}$       & $\sqrt{\frac{2(n+4)}{5}}$         & $ -\sqrt{\frac{n+1}{3}} $           &0               &0 \\
2 & 2  & $\sqrt{\frac{3(n+5)}{7}}$         & $\sqrt{\frac{n+5}{7}}$       & $\sqrt{\frac{3(n+5)}{105}}$       &0               &0             &$-\sqrt{\frac{2n}{5}}   $            \\
2 & 1  & $\sqrt{\frac{2(n+5)}{7}}$         & $2\sqrt{\frac{6(n+5)}{105}}$ & $3\sqrt{\frac{n+5}{105}}$         &0             &$\sqrt{\frac{n}{5}}$             &$-\sqrt{\frac{n}{5}}$            \\
2 & 0  & $3\sqrt{\frac{2(n+5)}{105}}$      & $3\sqrt{\frac{3(n+5)}{105}}$ & $3\sqrt{\frac{2(n+5)}{105}}$      &$-\sqrt{\frac{n}{15}}$               &$2\sqrt{\frac{n}{15}} $              &$-\sqrt{\frac{n}{15}}$                \\
2 & -1 & $3\sqrt{\frac{n+5}{105}}$         & $2\sqrt{\frac{6(n+5)}{105}}$ & $\sqrt{\frac{2(n+5)}{7}}$         & $-\sqrt{\frac{n}{5}}$              &$\sqrt{\frac{n}{5}}$               &0                \\
2 & -2 & $\sqrt{\frac{3(n+5)}{105}}$       & $\sqrt{\frac{n+5}{7}}$       & $\sqrt{\frac{3(n+5)}{7}}$         &$-\sqrt{\frac{2n}{5}}$               &0               &0                \\
3 & 3  & $\frac{2}{3}\sqrt{n+6}$           & $\frac{1}{3}\sqrt{n+6}$      & $\frac{1}{3}\sqrt{\frac{n+6}{7}}$ &0               &0               &$-\sqrt{\frac{3(n-1)}{7}}$               \\
3 & 2  & $\sqrt{\frac{n+6}{3}}$            & $2\sqrt{\frac{n+6}{21}}$     & $\sqrt{\frac{n+6}{21}}$           &0               &$\sqrt{\frac{n-1}{7}}$               &$-\sqrt{\frac{2(n-1)}{7}}$                \\
3 & 1  & $\sqrt{\frac{5(n+6)}{21}}$        & $\sqrt{\frac{5(n+6)}{21}}$   & $\sqrt{\frac{2(n+6)}{21}}$        &$-\sqrt{\frac{n-1}{35}}$               & $2\sqrt{\frac{2(n-1)}{35}}$              &$-\sqrt{\frac{6(n-1)}{35}}$                \\
3 & 0  & $\sqrt{\frac{10(n+6)}{63}}$       & $\sqrt{\frac{16(n+6)}{63}}$  & $\sqrt{\frac{10(n+6)}{63}}$       &$-\sqrt{\frac{3(n-1)}{35}}$               &$3\sqrt{\frac{n-1}{35}}$               &$-\sqrt{\frac{3(n-1)}{35}}$                \\
3 & -1 & $\sqrt{\frac{2(n+6)}{21}}$        & $\sqrt{\frac{5(n+6)}{21}}$   & $\sqrt{\frac{5(n+6)}{21}}$        &$-\sqrt{\frac{6(n-1)}{35}}$               &$2\sqrt{\frac{2(n-1)}{35}}$               & $-\sqrt{\frac{n-1}{35}}$               \\
3 & -2 & $\sqrt{\frac{n+6}{21}}$           & $2\sqrt{\frac{n+6}{21}}$     & $\sqrt{\frac{n+6}{3}}$            &$-\sqrt{\frac{2(n-1)}{7}}$               &$\sqrt{\frac{n-1}{7}}$               &0                \\
3 & -3 & $\frac{1}{3}\sqrt{\frac{n+6}{7}}$ & $\frac{1}{3}\sqrt{n+6}$      & $\frac{2}{3}\sqrt{n+6}$           &$-\sqrt{\frac{3(n-1)}{7}}$               &0               &0\\
 \cline{1-8}               
\end{tabular}\label{table}
\end{table*}
\section{CONSERVATION LAWS}\label{sec:cons_law}
In this Appendix, we derive the conservation laws associated with the ${\rm U}(1)$ gauge and spin rotational symmetries. We also derive the expressions for mass current (\ref{eq;j_m}) and spin current (\ref{eq;j_s}) from the GL action (\ref{act:gl}).\\ 
\indent We first study the conservation law associated with the ${\rm U}(1)$ gauge symmetry. The following argument is based on Refs.~\cite{sachdev11,polkovnikov05}. Under an infinitesimal ${\rm U}(1)$ gauge transformation $\Psi_{\alpha} \rightarrow e^{i\delta \theta(\bm{r})} \Psi_\alpha \sim \Psi_\alpha + i\delta \theta (\bm{r}) \Psi_{\alpha}$, the variation of the GL action (\ref{act:gl}) is given as
\begin{eqnarray}
\delta S^\mathrm{GL}
&=& i \delta \theta \int^\infty_0 {\rm d} \tau \int{\rm d}^dx 
\left\{
K\partial_\tau({\bf \Psi}^\dagger{\bf \Psi})\right.\nonumber \\
&&
+
J\partial_\tau[(\partial_\tau{\bf \Psi}^\dagger){\bf \Psi}
-
{\bf \Psi}^\dagger\partial_\tau{\bf \Psi}]\nonumber\\
&&
+
\left.\frac{1}{2m^*}\nabla\cdot[(\nabla {\bf \Psi}^\dagger){\bf \Psi}-{\bf \Psi}^\dagger\nabla{\bf \Psi}]\right\}.\label{dSgl}
\end{eqnarray}
From $\delta S^{\mathrm{GL}} = 0$ for arbitrary $\delta \theta$, we obtain the continuity equation
\begin{eqnarray}
\partial_t(K\rho+Jq)+\nabla\cdot\bm{j}_{\rm m}=0,\label{c-equation}
\end{eqnarray}
where $\rho$ and $\bm{j}_{\rm m}$ denote the superfluid density and superfluid mass current, respectively, defined as
\begin{eqnarray}
\rho
&=&
{\bf \Psi}^\dagger{\bf \Psi},\label{rho}\\
q
&=&
\frac{1}{i}\left[(\partial_t{\bf \Psi}^\dagger) {\bf \Psi}
-
{\bf \Psi}^\dagger(\partial_t{\bf \Psi})\right],\label{charge}\\
\bm{j}_{\rm m}
&=&
\frac{1}{2im^*}
\left[
{\bf \Psi}^\dagger\nabla{\bf \Psi}
-
(\nabla{\bf \Psi}^\dagger){\bf \Psi}
\right].\label{mass_j}
\end{eqnarray}
We clarify the physical meaning of $q$ below.
If $K\neq0$, neglecting $q$, Eq.\,(\ref{c-equation}) reduces to
\begin{eqnarray}
\partial_t\rho+\frac{1}{K}\nabla\cdot\bm{j}_{\rm m}=0.\label{c-equation-1}
\end{eqnarray}
Equation\,(\ref{c-equation-1}) represents the conservation of superfluid density. \\
\indent If $K=0$, keeping $q$, Eq.\,(\ref{c-equation-1}) reduces to 
\begin{eqnarray}
\partial_t q+\frac{1}{J}\nabla\cdot \bm{j}_{\rm m}=0.
\end{eqnarray}\\
\indent Before discussing the interpretation of $q$, we show that $K=0$ holds for a commensurate filling in the SF phase.
We first note that Eq.\,(\ref{eq:S}) is invariant under the following local gauge transformation 
\begin{eqnarray}
\begin{split}
b_{i\alpha} &\rightarrow e^{i\eta(\tau)}b_{i\alpha},\\
\phi_{i\alpha} &\rightarrow e^{i\eta(\tau)}\phi_{i\alpha},\\
\mu &\rightarrow \mu+i\partial_\tau\eta(\tau).\label{gauge_}
\end{split}
\end{eqnarray}
The GL action (\ref{act:gl}) should be invariant under the corresponding transformation $\Psi_\alpha \rightarrow e^{i \eta(\tau)}\Psi_\alpha$, which requires the condition
\begin{eqnarray}
K=-\frac{\partial C_2}{\partial\mu}.\label{nui}
\end{eqnarray}
We can directly confirm Eq.\,(\ref{nui}) from Eqs.\,(\ref{C-2}) and (\ref{K-}). On the other hand, in the vicinity of the phase boundary, the ground-state energy can be expanded by the superfluid density $n_{\rm c}$ as
\begin{eqnarray}
E=E^0(0,n)+C_2n_{\rm c}+O(n_{\rm c}^2).
\end{eqnarray}
The expectation value of the filling number is given by
\begin{eqnarray}
\langle \hat n \rangle=-\frac{\partial E}{\partial \mu}\sim n - \frac{\partial C_2}{\partial \mu}n_{\rm c}.
\end{eqnarray}
In the MI phase, $\langle \hat{n} \rangle$ is an integer because $n_{\rm c}=0$ as expected.
Meanwhile, since $n_{\rm c}\neq 0$, $\langle \hat{n} \rangle$ takes an integer value only if $-\partial C_2/\partial \mu=0$ in the SF phase. Thus, $K=0$ holds for a commensurate filling in the SF phase. Furthermore, when $K=0$, the TDGL equation (\ref{tdgl}) is invariant under the transformation $\Psi_\alpha \leftrightarrow \Psi_\alpha^*$, i.e., particle-hole symmetric
\cite{polkovnikov05,fisher89,nakayama15}.\\
\indent Equation (\ref{gauge_}) implies that the transformation $b_{i\alpha}\rightarrow e^{i\eta}b_{i\alpha}$ and $\phi_{i\alpha}\rightarrow e^{i\eta}\phi_{i\alpha}$ is equivalent to the shift of $\mu$: $\mu \rightarrow \mu-i\partial_\tau \eta$ \cite{fisher89}.
The infinitesimal shift of the chemical potential $\mu \rightarrow \mu+i\partial \delta \eta$ is, therefore, equivalent to the transformation of $\Psi_\alpha$: 
\begin{eqnarray}
\Psi_\alpha \rightarrow e^{-i\delta \eta (\tau)} \Psi_\alpha \sim \Psi_\alpha -i \delta \eta(\tau) \Psi_\alpha. \label{eq:Psi}
\end{eqnarray}
The variation of $S^{\rm GL}$ under Eq.\,(\ref{eq:Psi}) is given by
\begin{eqnarray}
\delta S^\mathrm{GL} = -\delta \mu \int {\rm d}^dx (K \rho + Jq), \label{dact:gl}
\end{eqnarray} 
where $\delta \mu = i\partial_\tau \eta (\tau)$. Comparing Eq.\,(\ref{dact:gl}) with $\delta S^\mathrm{GL} = (\delta S^\mathrm{GL}/\delta \mu(\tau)) \delta \mu(\tau)$, we obtain
\begin{eqnarray} 
-\frac{\delta S^\mathrm{GL}}{\delta \mu(\tau)} = \int {\rm d}^d x (K \rho + J q). \label{eq:S/mu}
\end{eqnarray}
The left-hand side of Eq.\,(\ref{eq:S/mu}) represents the deviation of particle number from the static value.
For a commensurate filling, setting $K=0$, Eq.\,(\ref{eq:S/mu}) reduces to
\begin{eqnarray}
-\frac{\delta S^\mathrm{GL}}{\delta \mu (\tau)}= \int {\rm d}^dx Jq.
\end{eqnarray}
We thus find that $q$ means the deviation of particle density from a commensurate filling.\\
\indent We next derive the conservation law associated with the spin rotational symmetry. We introduce an infinitesimal spin rotation about a unit vector $\bm{n}$:
\begin{eqnarray}
{\bf \Psi} \rightarrow e^{-i(\bm{n} \cdot\bm{F})\delta \lambda} {\bf \Psi} 
\sim
{\bf \Psi} - i(\bm{n} \cdot\bm{F})\delta \lambda {\bf \Psi}. \label{spin_rotation}
\end{eqnarray}  
The variation of the GL action\,(\ref{act:gl}) under Eq.\,(\ref{spin_rotation}) is given by
\begin{eqnarray}
\delta S^\mathrm{GL}
&=&
-i\delta\lambda  \int^\infty_0 {\rm d} \tau  \int {\rm d}^dx
\left\{
K\partial_\tau
\left[
 {\bf \Psi}^\dagger ( \bm{n}\cdot\bm{F}){\bf \Psi}
 \right] 
 \right.\nonumber\\
 &+&
 J\partial_\tau
 \left[
 (\partial_\tau{\bf \Psi}^\dagger)( \bm{n}\cdot\bm{F}){\bf \Psi}
 -
 {\bf \Psi}^\dagger( \bm{n}\cdot\bm{F}) \partial_\tau{\bf \Psi}
 \right]\nonumber\\
 &+&
 \left.
 \frac{1}{2m^*}\nabla\cdot
 \left[
 (\nabla{\bf \Psi}^\dagger)( \bm{n}\cdot\bm{F}){\bf \Psi}
 -
 {\bf \Psi}^\dagger( \bm{n}\cdot\bm{F})\nabla{\bf \Psi}
 \right]
 \right\}.\nonumber \\
\end{eqnarray}
$\delta S^\mathrm{GL} = 0$ for arbitrary $\delta \lambda $ yields the continuity equation
\begin{eqnarray}
\partial_t(K \rho_{\rm s}^{\bm{n}} + J q_{\rm s}^{\bm{n}}) + \nabla\cdot \bm{j}_{\rm s}^{\bm{n}}=0,\label{s-c-eq}
\end{eqnarray}
where $\rho_{\rm s}^{\bm{n}}$ denotes the magnetization density for the component along $\bm{n}$ and $\bm{j}_{\rm s}^{\bm{n}}$ the spin current for $\rho_{\rm s}^{\bm{n}}$. They are defined  as 
\begin{eqnarray}
\rho_{\rm s}^{\bm{n}} 
&=&
{\bf \Psi}^\dagger  (\bm{n} \cdot \bm{F})  {\bf \Psi},\\
q_{\rm s}^{\bm{n}} 
&=&
\frac{1}{i} 
\left[
 (\partial_t {\bf \Psi}^\dagger)(\bm{n} \cdot \bm{F}) {\bf \Psi}
 -
 {\bf \Psi}^\dagger(\bm{n} \cdot \bm{F}) \partial_t{\bf \Psi})
 \right],\\
\bm{j}_{\rm s}^{\bm{n}} 
&=&
\frac{1}{2im^*}
\left[
{\bf \Psi}^\dagger (\bm{n}\cdot\bm{F})\nabla{\bf \Psi}
-
(\nabla{\bf \Psi}^\dagger)(\bm{n}\cdot\bm{F}){\bf \Psi}
\right].\nonumber \\ \label{spin_j}
\end{eqnarray}
We clarify the physical meaning of $q^{\bm{n}}_{\rm s}$ below.
If $K \neq 0$, neglecting $q^{\bm n}_{\rm s}$, Eq.\,(\ref{s-c-eq}) represents the conservation of magnetization for the component along $\bm{n}$. \\
\indent Analogous to Eq.\,(\ref{gauge_}), we consider the following trans formation under a spin rotation
\begin{eqnarray}
\begin{split}
\vec{b}_i &\rightarrow e^{-i ( \bm{n} \cdot \bm{F}  )\nu(\tau)} \vec{b}_i,\\
\vec{\phi}_i &\rightarrow e^{-i (\bm{n} \cdot \bm{F} )  \nu (\tau)} \vec{\phi}_i,
\end{split}\label{tr:spin_t}
\end{eqnarray}
where $\vec{b}_i =(b_{i1},b_{i0},b_{i-1})^T$ and $\vec{\phi}_i =  (\phi_{i1},\phi_{i0},\phi_{i-1})^T$.
Under the transformation (\ref{tr:spin_t}), Eq.\,(\ref{eq:S}) is transformed as
\begin{eqnarray}
S \rightarrow S - \int^\beta_0 {\rm d} \tau \bm{H} (\tau) \cdot  \int {\rm d}^d x \langle \bm{F} \rangle_b. \label{L -> L-HS}
\end{eqnarray}
Here, $\bm{H} = i\partial_\tau \nu \bm{n}$ and $ \langle \bm{F} \rangle_b=\sum_{\alpha,\beta}b_{i\alpha}^* (e^{i\bm{F}\cdot \bm{n} \nu} \bm{F} e^{-i\bm{F}\cdot\bm{n}\nu})_{\alpha\beta}b_{i\beta}$.
Equation\,(\ref{L -> L-HS}) shows that the transformation (\ref{tr:spin_t}) is equivalent to application of magnetic field $\bm{H}$ in the direction of $\bm{n}$.

The transformation of ${\bf \Psi}$ that corresponds to Eq.\,(\ref{tr:spin_t}) is ${\bf \Psi}\rightarrow e^{-i(\bm{n}\cdot \bm{F}) \nu(\tau)} {\bf \Psi}$.
Under an infinitesimal spin rotation ${\bf \Psi} \rightarrow e^{-i ( \bm{n} \cdot \bm{F})\delta \nu}{\bf \Psi}\sim {\bf \Psi}-i ( \bm{n}\cdot \bm{F} )\delta \nu {\bf \Psi}$, the variation of Eq.\,(\ref{act:gl}) is given by
\begin{eqnarray}
\delta S^\mathrm{GL} = -\delta \bm{H} (\tau) \cdot \int {\rm d}^d x( K \langle \! \langle \bm{F} \rangle \! \rangle + J \bm{q}_{\rm s} ),\label{Lgl -> Lgl -HS}
\end{eqnarray}
where $\delta \bm H(\tau) = i\delta \partial_\tau \nu (\tau) \bm n$. Comparing Eq.\,(\ref{Lgl -> Lgl -HS}) and
\begin{eqnarray}
\delta S^\mathrm{GL}= \frac{\delta S^\mathrm{GL}}{\delta \bm H} \cdot \delta \bm H,
\end{eqnarray}
we obtain
\begin{eqnarray}
- \frac{\delta S^\mathrm{GL}}{\delta \bm{H}} = \int {\rm d}^dx ( K \langle \! \langle \bm{F} \rangle \! \rangle + J \bm{q}_{\rm s} ),\label{eq:dS/dH}
\end{eqnarray}
where $\bm{n}\cdot\bm{q}_{\rm s}=q_{\rm s}^{\bm{n}}$.
The left-hand side of Eq.\,(\ref{eq:dS/dH}) represents the magnetization along $\bm{H}$. In the case of a commensurate filling, i.e., $K=0$, we obtain
\begin{eqnarray}
- \frac{\delta S^\mathrm{GL}}{\delta \bm{H}} = \int {\rm d}^d x J \bm{q}_{\rm s} .\label{eq:dS/dH_K=0}
\end{eqnarray}
We thus find that the $q_{\rm s}^{\bm{n}}$ represents the deviation of magnetization density from the value for a commensurate filling.
\bibliography{critical_v.bib} 

\begin{thebibliography}{63}
\bibitem [{hea()}]{healing_length}%
  \BibitemOpen
  \href@noop {} {}\bibinfo {note} {The healing length $\xi$ can be derived as
  follows: substituting ${\bf \Psi} = \sqrt{\rho^{\rm 0} + \delta
  \rho(x)}(0,1,0)^T$ into the TDGL equation (\ref{TDGL}) and linearizing by
  $\delta \rho(x)$, one obtains $\delta \rho(x) \propto e^{-x/\xi}$, where
  $\rho^{\rm 0}$ is the equilibrium SF density and $\delta \rho(\infty)=0$ is
  assumed.}\BibitemShut {Stop}%
\end{thebibliography}
\end{document}